\documentclass[12pt,aps,prb,preprint,showpacs,showkeys]{revtex4}

\usepackage{graphicx}
\usepackage{subfigure}
\usepackage{amsmath}
\usepackage{amssymb}

\begin{document}

\title{Frustrated mixed spin-1/2 and spin-1 Ising ferrimagnets on a triangular lattice}
\author{M. \v{Z}ukovi\v{c}}
 \email{milan.zukovic@upjs.sk}
\author{A. Bob\'ak}
 \affiliation{Department of Theoretical Physics and Astrophysics, Faculty of Science,\\ 
P. J. \v{S}af\'arik University, Park Angelinum 9, 041 54 Ko\v{s}ice, Slovakia}
\date{\today}

\begin{abstract}
Mixed spin-1/2 and spin-1 Ising ferrimagnets on a triangular lattice with sublattices A, B and C are studied for two spin value distributions $(S_{\rm A},S_{\rm B},S_{\rm C})=(1/2,1/2,1)$ and $(1/2,1,1)$ by Monte Carlo simulations. The non-bipartite character of the lattice induces geometrical frustration in both systems, which leads to the critical behavior rather different from their ferromagnetic counterparts. We confirm second-order phase transitions belonging to the standard Ising universality class occurring at higher temperatures, however, in both models these change at tricritical points (TCP) to first-order transitions at lower temperatures. In the model $(1/2,1/2,1)$, TCP occurs on the boundary between paramagnetic and ferrimagnetic $(\pm 1/2,\pm 1/2,\mp 1)$ phases. The boundary between two ferrimagnetic phases $(\pm 1/2,\pm 1/2,\mp 1)$ and $(\pm 1/2,\mp 1/2,0)$ at lower temperatures is always first order and it is joined by a line of second-order phase transitions between the paramagnetic and the ferrimagnetic $(\pm 1/2,\mp 1/2,0)$ phases at a critical endpoint. The tricritical behavior is also confirmed in the model $(1/2,1,1)$ on the boundary between the paramagnetic and ferrimagnetic $(0,\pm 1,\mp 1)$ phases.
\end{abstract}

\pacs{05.50.+q, 64.60.De, 75.10.Hk, 75.30.Kz, 75.50.Gg}

\keywords{Mixed-spin system, Frustrated Ising ferrimagnet, Triangular lattice, Monte Carlo simulation, Tricritical point, Critical endpoint}



\maketitle

\section{Introduction}
\hspace*{5mm} Mixed-spin Ising systems have been mostly investigated as possible models of some types of ferrimagnetic and molecular-based magnetic materials. The used approaches include an exact treatment in special cases~\cite{gonc85,lipo95,jasc98,dakh98,jasc05}, mean-field approximation~\cite{kane91,abub01}, effective-field theory with correlations~\cite{kane87,boba97,boba98,kane98a,kane98b,boba00,boba02}, Monte Carlo simulations~\cite{zhan93,buen97,selk00,naka00,naka02,oitm03,godo04,selk10} and some other methods~\cite{iwas84,vero88,tuck01,godo04a,oitm05,oitm06}. The main focus were their phase diagrams as well as technologically interesting compensation behavior with possibility to achieve zero total magnetization by tuning of temperature below the critical point. Most of the studies considered the simplest models consisting of two sublattices one of which is occupied with spins $S=1/2$ and the other with $S=1$. Such a mixed-spin model can be described by the Hamiltonian
\begin{equation}
\label{Hamiltonian}
H=-J\sum_{\langle i,j \rangle}\sigma_{i}S_{j}-D\sum_{j}S_{j}^2,
\end{equation}
where $\sigma_{i}=\pm 1/2$ and $S_{j}=\pm 1,0$ are spins on different sublattices, $\langle i,j \rangle$ denotes the sum over nearest neighbors, $J<0$ is a antiferromagnetic exchange interaction parameter and $D$ is a single-ion anisotropy parameter. Negative values of the parameter $D$ favor nonmagnetic states with $S_{j}=0$ and positive values magnetic states with $S_{j}=\pm 1$.

Due to persisting ambiguities majority of the investigations focused on the simplest lattices, i.e., the square in two and cubic in three dimensions. We note that a long standing controversy regarding the critical and compensation behaviors even for the most studied case of the model on a square lattice was solved only recently by Monte Carlo simulation that has convincingly shown~\cite{selk10} that there are neither tricritical nor compensation points, as had been suggested by some previous approximative approaches~\cite{kane87,kane91,boba97,oitm06}. On the other hand, in the same study the presence of both the tricritical point and a line of compensation points was confirmed in the three-dimensional model on a simple cubic lattice. This finding might suggest that the increased dimensionality is responsible for the appearance of the tricritical and compensation behaviors. Nevertheless, our recent study on a triangular lattice ferromagnet~\cite{zuko15} demonstrated that the tricritical point can also appear in a two-dimensional lattice as long as the coordination number is sufficiently high. The effect of the coordination number in the presence of bond disorder on tricritical behavior of a two dimensional system was also recently studied in a random Blume-Capel model on a triangular lattice~\cite{theo12}.

\begin{figure}[t!]
\centering
\subfigure{\includegraphics[scale=0.57,clip]{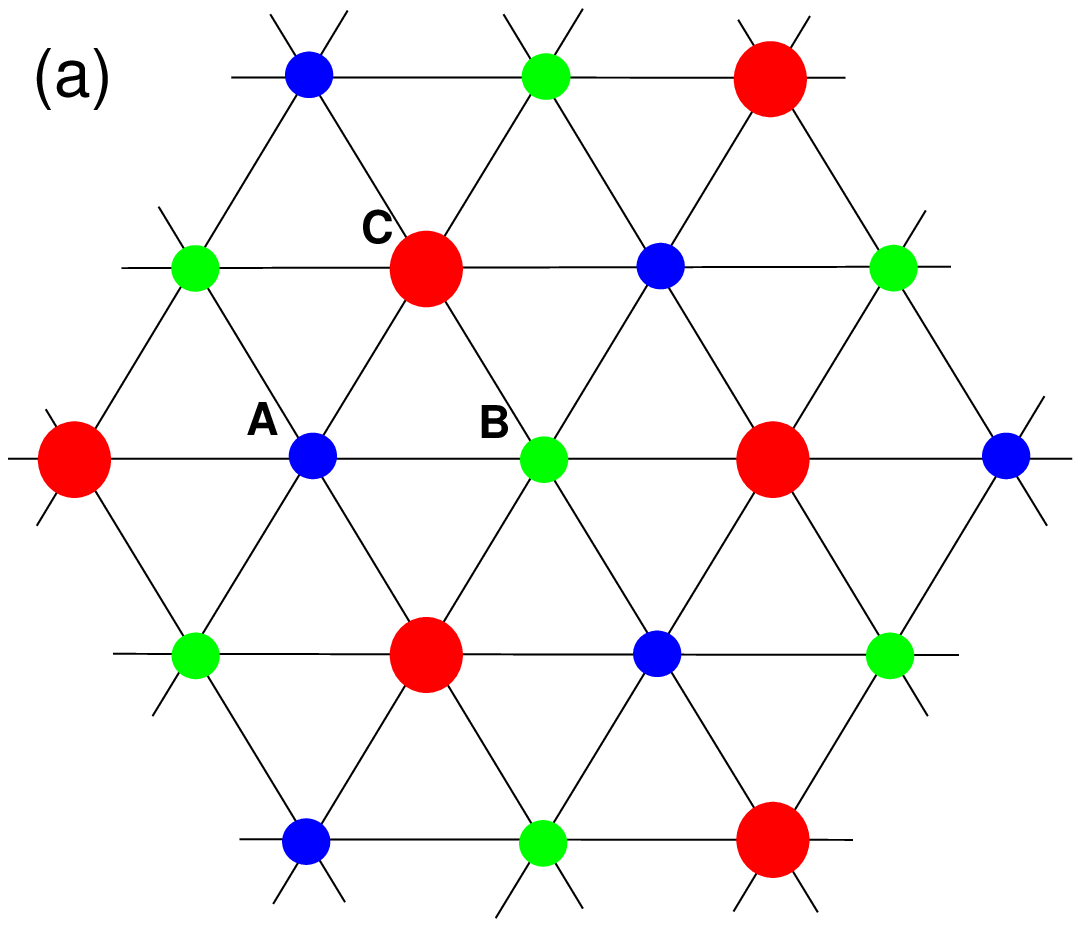}\label{fig:schem_a}}\hspace*{5mm}
\subfigure{\includegraphics[scale=0.57,clip]{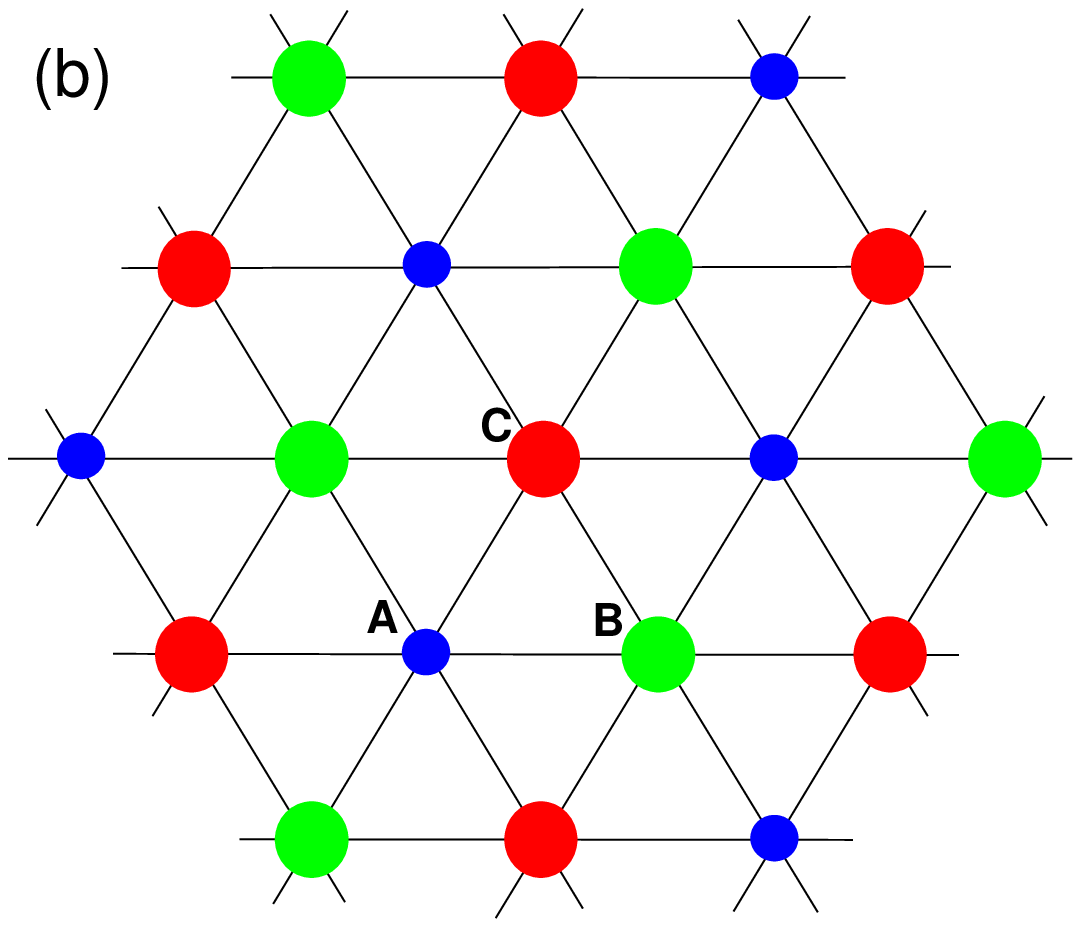}\label{fig:schem_b}}
\caption{(Color online) Mixed-spin ${\bf S}=(S_{\rm A},S_{\rm B},S_{\rm C})$ models on a triangular lattice consisting of sublattices A, B and C, with (a) ${\bf S}=(1/2,1/2,1)$ mixing - model I and (b) ${\bf S}=(1/2,1,1)$ mixing - model II. Small and large circles denote spin-1/2 and spin-1 sites, respectively.}\label{fig:models}
\end{figure} 

We point out that the previous studies were performed on bipartite lattices, in which case the sign of the exchange interaction is irrelevant to the thermodynamic and critical properties of the model in the absence of an external field. On the other hand, the present mixed-spin model is considered on a non-bipartite triangular lattice, in which case the sign of the exchange interaction matters. Namely, in contrast to the ferromagnetic case, the ferrimagnetic interaction will induce geometrical frustration, which can be expected to have some impact on the critical behavior. As shown in Fig.~\ref{fig:models}, the lattice consists of three sublattices A, B and C, occupied with spins ${\bf S}=(S_{\rm A},S_{\rm B},S_{\rm C})$. This allows to further study the model in two different mixing modes. We can consider a mixed-spin ${\bf S}=(1/2,1/2,1)$ model I, as schematically depicted in Fig.~\ref{fig:schem_a}, in which one sublattice is occupied with spin $S=1$ sites and the remaining two sublattices with spin $S=1/2$ sites. Thus, each spin-1 site is surrounded by $z=6$ nearest neighbors with spin $S=1/2$. The other way of the spin-mixing is realized in a ${\bf S}=(1/2,1,1)$ model II, shown in Fig.~\ref{fig:schem_b}, which is obtained when the spin-1/2 and spin-1 sites in the model I are swapped. The two models were shown to display qualitatively different critical behaviors even for the ferromagnetic exchange interactions~\cite{zuko15}.

The goal of the present study is to examine effects of the geometrical frustration on the critical behavior of the above defined ferrimagnetic mixed-spin systems, to determine their phase diagrams and to confront them with their ferromagnetic counterparts as well as the pure spin-1/2 and spin-1 antiferromagnetic systems.

\section{Monte Carlo simulation}
\begin{figure}[t]
\centering
    \includegraphics[scale=0.57]{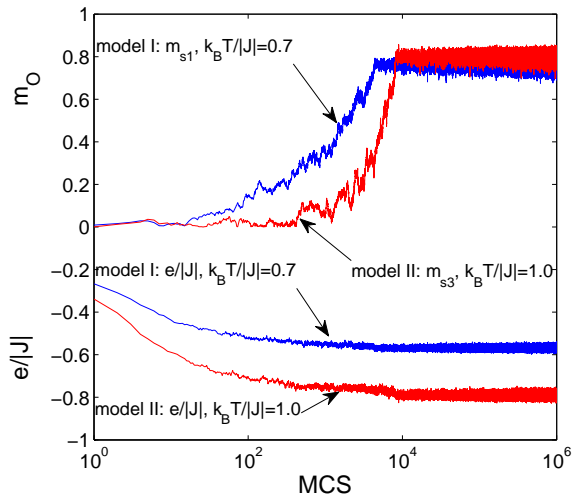}
\caption{(Color online) Time evolutions of the internal energy per site $e/|J|$ and the staggered magnetizations $m_{s1},m_{s3}$ (see definitions below), starting from the disordered phase at the temperatures $k_BT/|J|=0.7$ in model I and 1.0 in model II, for $D/|J|=0$ and $L=120$.}\label{fig:noneq}
\end{figure}

In order to study the behavior of various thermodynamic quantities in the parameter space and to determine the phase diagrams we use Monte Carlo (MC) simulations with the Metropolis update rule and employ the periodic boundary conditions. We consider lattices with the size $L \times L$, with $L$ ranging from $24$ up to $120$. We perform $N=2\times 10^5$ up to $10^6$ MCS (Monte Carlo sweeps), the first $20\%$ of which are used to bring the system to equilibrium and then discarded, and the remaining data are used to estimate thermal averages and statistical errors. In order to demonstrate that the used MCS is sufficient to ensure equilibrium conditions, in Fig.~\ref{fig:noneq} we present MC time evolutions of some relevant quantities, such as the order parameters $m_{s1}$ (model I) and $m_{s3}$ (model II) and the internal energy per site $e/|J|$, starting from the disordered phase. The simulations are performed in both models for the largest of the considered lattice sizes $L=120$, which is the most difficult to equilibrate, at the temperatures in the vicinity of the respective critical points for $D/|J| = 0$. The plots show that in these cases the equilibrium is reached in less than $10^4$ MCS. 

The phase boundaries are roughly determined from the maxima of some thermodynamic functions, such as the specific heat, for a selected fixed value of $L$. We chose $L=48$, as a compromise value above which the specific heat maxima positions do not change considerably and the phase diagrams can be determined in a relatively wide parameter space in a reasonable computational time. In the region where the critical line as a function of the single-ion anisotropy parameter $D$ is more or less horizontal it is convenient to obtain temperature dependencies of the calculated quantities at a fixed value of $D$. In such a case simulations start from the paramagnetic phase using random initial configurations with the temperature gradually decreased and a new simulation starting from the final configuration obtained at the previous temperature. On the other hand, if the phase boundary shape changes to vertical we obtain variations of the quantities as functions of the single-ion anisotropy parameter $D$ at a fixed temperature. Then simulations start from appropriately chosen states (i.e., not necessarily random), expected in the considered region of the parameter space. Following the above described approach we ensure that the system is maintained close to the equilibrium in the entire range of the changing parameter and thus considerably shortens thermalization periods. In order to estimate statistical errors, we perform three independent simulations at all considered parameter values. 

At some selected points of the phase boundaries we perform a more thorough finite-size scaling (FSS) analysis in order to determine more precisely the location of the critical points and the corresponding critical exponents. In such a case we perform more extensive simulations using up to $N=10^7$ MCS and apply the reweighing techniques~\cite{ferr88}. For more reliable estimation of statistical errors, in this case we used the $\Gamma$-method~\cite{wolf04}. Having obtained the maxima of the relevant quantities, we apply the linear fitting procedure for logarithms of data with errors, following the method in York et al.~\cite{york04}. In order to assess the quality of the fitting, as a measure of goodness of fit we evaluated an adjusted coefficient of determination~\cite{theil61} of the linear fit $R^2$. The critical points and the exponents are then extracted from the FSS analysis, using the linear sizes $L=24,48,72,96$ and $120$. 

We calculate the following quantities: the internal energy per spin $e=\langle H \rangle/L^2$, the respective sublattice magnetizations per site $m_{\rm X}$, (X = A, B or C), as order parameters on the respective sublattices, which for the model I are given by
\begin{equation}
\label{subAB_mag}
m_{\rm A(B)} = 3\langle |M_{\rm A(B)}| \rangle/L^2 = 3\Big\langle\Big|\sum_{i \in {\rm A(B)}}\sigma_{i}\Big|\Big\rangle/L^2,
\end{equation}
\begin{equation}
\label{subC_mag}
m_{\rm C} = 3\langle |M_{\rm C}| \rangle/L^2 = 3\Big\langle\Big|\sum_{i \in {\rm C}}S_{i}\Big|\Big\rangle/L^2,
\end{equation}
and for the model II by
\begin{equation}
\label{subA_mag}
m_{\rm A} = 3\langle |M_{\rm A}| \rangle/L^2 = 3\Big\langle\Big|\sum_{i \in {\rm A}}\sigma_{i}\Big|\Big\rangle/L^2,
\end{equation}
\begin{equation}
\label{subBC_mag}
m_{\rm B(C)} = 3\langle |M_{\rm B(C)}| \rangle/L^2 = 3\Big\langle\Big|\sum_{i \in {\rm B(C)}}S_{i}\Big|\Big\rangle/L^2,
\end{equation}
where $\langle\cdots\rangle$ denotes thermal average. Based on the ground-state considerations (see below), for the identified ordered phases we additionally define the following order parameters for the entire system, which take values between $0$ in the fully disordered and $1$ in the fully ordered phase. For the model I we introduce two order parameters (staggered magnetizations per site) $m_{s1}$ and $m_{s2}$ given by
\begin{equation}
\label{m_s1}
m_{s1} = \langle |M_{s1}| \rangle/L^2 = \Big\langle\Big|2\sum_{i \in {\rm A}}\sigma_{i}+2\sum_{j \in {\rm B}}\sigma_{j}-\sum_{k \in {\rm C}}S_{k}\Big|\Big\rangle/L^2,
\end{equation}
and 
\begin{equation}
\label{m_s2}
m_{s2} = \langle |M_{s2}| \rangle/L^2 = 3\Big\langle\Big|\sum_{i \in {\rm A}}\sigma_{i}-\sum_{j \in {\rm B}}\sigma_{j}\Big|\Big\rangle/L^2.
\end{equation}
For the model II we define the order parameter $m_{s3}$ as
\begin{equation}
\label{ms3}
m_{s3} = \langle |M_{s3}| \rangle/L^2 = 3\Big\langle\Big|\sum_{j \in {\rm B}}S_{j}-\sum_{k \in {\rm C}}S_{k}\Big|\Big\rangle/2L^2.
\end{equation}
Unlike in ferrimagnetic systems on bipartite lattices, the three sublattices in the present system facilitate spin arrangements in such a way that the total net magnetization is always zero and, therefore, of no practical use. 

Further, we calculate the susceptibilities pertaining to the respective order parameters $O=M_{\rm X}$ (X = A, B, C) and also $O=M_{si}$ ($i=1,2$ or $3$)
\begin{equation}
\label{chi}\chi_{O} = \frac{\langle O^{2} \rangle - \langle O \rangle^{2}}{N_Ok_{B}T}, 
\end{equation}
the specific heat per site $c$
\begin{equation}
\label{c}c=\frac{\langle H^{2} \rangle - \langle H \rangle^{2}}{N_Ok_{B}T^{2}},
\end{equation}
where $N_O$ is the number of sites on the (sub)lattice on which $O$ is defined. Further, we define the logarithmic derivatives of $\langle O \rangle$ and $\langle O^{2} \rangle$ with respect to $\beta=1/k_{B}T$,
\begin{equation}
\label{D1}D1_{O} = \frac{\partial}{\partial \beta}\ln\langle O \rangle = \frac{\langle O H
\rangle}{\langle O \rangle}- \langle H \rangle,
\end{equation}
\begin{equation}
\label{D2}D2_{O} = \frac{\partial}{\partial \beta}\ln\langle O^{2} \rangle = \frac{\langle O^{2} H
\rangle}{\langle O^{2} \rangle}- \langle H \rangle,
\end{equation}
and finally the fourth-order Binder cumulant $U_O$ corresponding to the quantity $O$
\begin{equation}
\label{U}U_{O} = 1-\frac{\langle O^{4}\rangle}{3\langle O^{2}\rangle^{2}} \ .
\end{equation}

The above standard thermodynamic quantities (\ref{chi}-\ref{c}), as well as the less traditional ones defined by Eqs.~(\ref{D1}-\ref{U}), serve to obtain estimates of the respective critical exponents by FSS analysis~\cite{ferr91,land00}. In particular, we use the following scaling relations, applied to the maximum values of the following functions:
\begin{equation}
\label{scalchi}\chi_{O,max}(L) \propto L^{\gamma_O/\nu_O},
\end{equation}
\begin{equation}
\label{scalc}c_{max}(L) \propto L^{\alpha/\nu_O},
\end{equation}
\begin{equation}
\label{scalD1}D1_{O,max}(L) \propto L^{1/\nu_O},
\end{equation}
\begin{equation}
\label{scalD2}D2_{O,max}(L) \propto L^{1/\nu_O},
\end{equation}
\noindent where $\alpha$ is the critical exponent of the specific heat and $\nu_O$, $\gamma_O$ are the critical exponents of the correlation length and susceptibility, respectively, pertaining to the quantity $O$. In case of a first-order phase transition, the above quantities~(\ref{scalchi}-\ref{scalD2}) are expected to scale as $\propto L^d$, where $d=2$ is the system dimension. The order parameter cumulant, defined by Eq.~(\ref{U}), can also serve for a simple yet relatively precise location of the phase transition point as a point at which the cumulant curves obtained for different system sizes intersect at a universal value, e.g., $U_O(T_c)=0.611$ for a two-dimensional Ising model~\cite{kame93}.

\section{Results}
\subsection{Ground state}
Let us first identify all the possible ground states (GS) for entire range of the single-ion anisotropy parameter $D$. Considering the lattice system consisting of three interpenetrating sublattices A, B and C, as schematically depicted in Fig.~\ref{fig:models}, the Hamiltonians of the respective models I and II can be defined as
\begin{equation}
\label{Hamiltonian_I}
H_{\rm I}=-J\Big(\sum_{i \in A,j \in B}\sigma_{i}\sigma_{j}+\sum_{i \in A,k \in C}\sigma_{i}S_{k}+\sum_{j \in B,k \in C}\sigma_{j}S_{k}\Big)-D\sum_{k \in C}S_{k}^2,
\end{equation}
\begin{equation}
\label{Hamiltonian_II}
H_{\rm II}=-J\Big(\sum_{i \in A,j \in B}\sigma_{i}S_{j}+\sum_{i \in A,k \in C}\sigma_{i}S_{k}+\sum_{j \in B,k \in C}S_{j}S_{k}\Big)-D\Big(\sum_{j \in B}S_{j}^2+\sum_{k \in C}S_{k}^2\Big).
\end{equation}

\begin{table}[t!]
\caption{Ground state configurations and the respective energies for different ranges of the single-ion anisotropy parameter.}
\label{tab:GS}
	\centering
		\begin{tabular}{ccccc}
		\hline
		Model & \multicolumn{2}{c}{I}  & \multicolumn{2}{c}{II} \\
		\hline
		$D/|J|$ & State & Energy $e/|J|$ & State & Energy $e/|J|$ \\
		\hline
		$(-\infty,-3/2)$ & ${\rm FR}_2^{\rm I}$: $(\pm 1/2,\mp 1/2,0)$ & $-1/4$ & P: $(0,0,0)$ & $0$ \\
		$(-3/2,\infty)$ & ${\rm FR}_1^{\rm I}$: $(\pm 1/2,\pm 1/2,\mp 1)$ & $-3/4-D/3|J|$ & ${\rm FR}^{\rm II}$: $(0,\pm 1,\mp 1)$ & $-1-2D/3|J|$ \\
		\hline
		\end{tabular}
\end{table}

Focusing on a triangular elementary unit cell consisting of the spins $S_{\rm A}$, $S_{\rm B}$, $S_{\rm C}$, from the Hamiltonians~(\ref{Hamiltonian_I}) and (\ref{Hamiltonian_II}) one can obtain expressions for the reduced energies per spin $e/|J|$ of different spin arrangements as functions of $D/|J|$. Then the ground states are determined as configurations corresponding to the lowest energies for different values of $D/|J|$, as tabulated in Table~\ref{tab:GS}. There are two long-range order (LRO) ferrimagnetic (FR) states ${\rm FR}_1^{\rm I}$ and ${\rm FR}_2^{\rm I}$ in the model I and one LRO ferrimagnetic state ${\rm FR}^{\rm II}$ and one disordered paramagnetic (P) phase in the model II. We note that while in ${\rm FR}_2^{\rm I}$ zero means nonmagnetic states ($S_{j}=0$) of spins on sublattice C, in ${\rm FR}^{\rm II}$ zero means magnetic states ($\sigma_{i}=\pm 1/2$) of spins on sublattice A but equally in states $+1/2$ and $-1/2$, thus giving zero net sublattice magnetization. The critical value of the single-ion anisotropy parameter separating the respective phases is the same for both models  $D_c/|J|=-3/2$.

\subsection{Monte Carlo}
\subsubsection{Model I: ${\bf S}=(1/2,1/2,1)$}

\begin{figure}[t!]
\centering
\subfigure{\includegraphics[scale=0.57,clip]{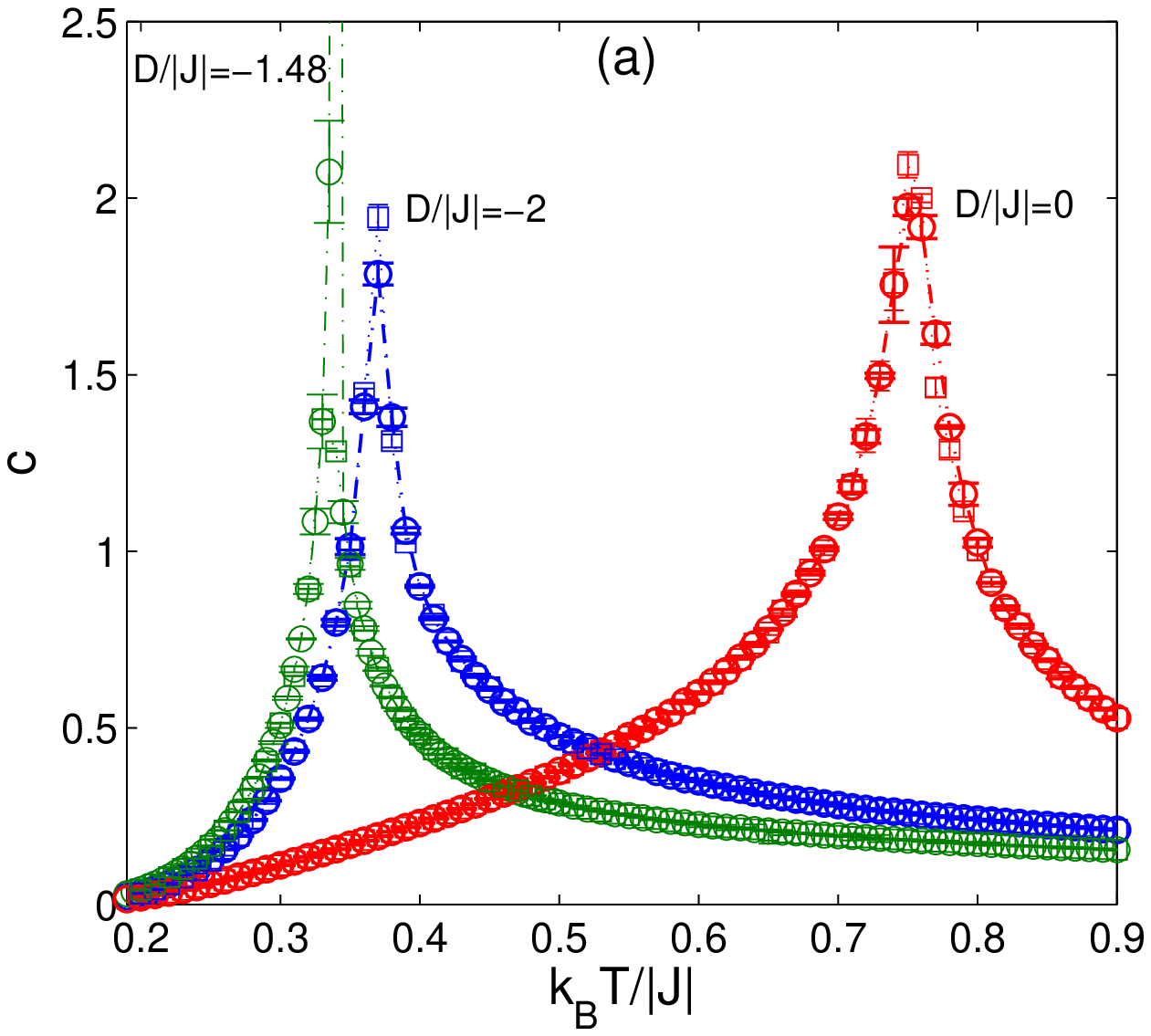}\label{fig:c-T_L48}}
\subfigure{\includegraphics[scale=0.57,clip]{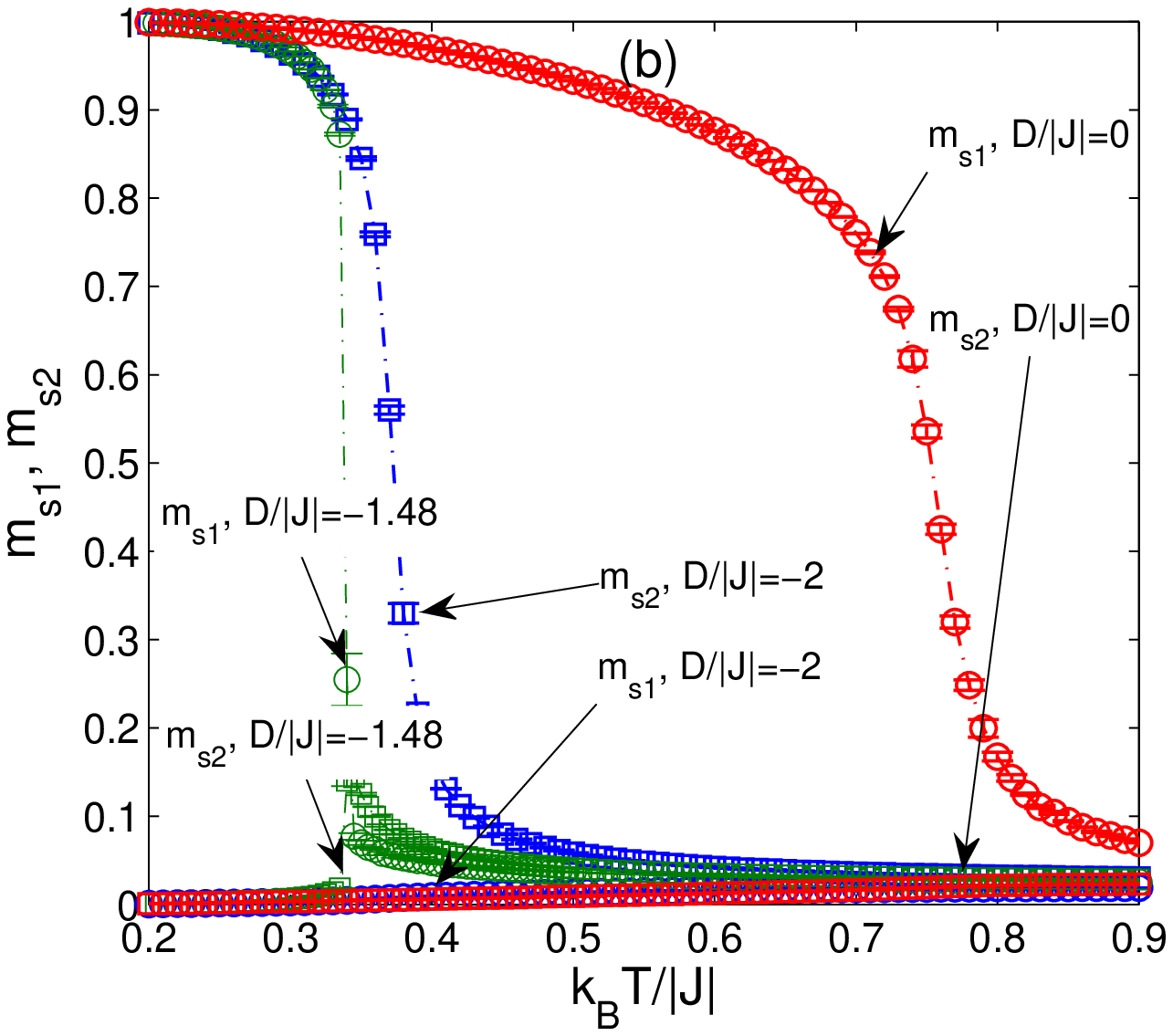}\label{fig:ms1_2-T_L48}}
\caption{(Color online) Temperature dependencies of (a) the specific heat and (b) the order parameters, for selected values of $D/|J|$ and a fixed lattice size $L=48$. In (a) additionally the results for $L=72$ are shown (squares).}\label{fig:x-T}
\end{figure}

Phase boundaries between the disordered paramagnetic and the respective ordered ferrimagnetic phases are determined from the specific heat maxima for a fixed $L=48$\footnote{Hence, in fact these are only pseudo-critical points} and the nature of the ordered phases is established from the introduced order parameters $m_{s1}$ and $m_{s2}$. These are shown in Figs.~\ref{fig:c-T_L48} and~\ref{fig:ms1_2-T_L48} for selected values of the parameter $D/|J|$ below, close to and above the critical value $D_c/|J|$. All the specific heat curves show pronounced sharp peaks, signifying phase transitions to the low-temperature ferrimagnetic states. However, the (pseudo)transition temperatures appear to be a nonmonotonic functions of $D/|J|$. More specifically, the transition temperature for $D/|J|=-1.48$, i.e., close to the critical value $D_c/|J|$, is lower than for the other two values below and above $D_c/|J|$. 
Moreover, the corresponding specific heat maximum has a spike-like shape and its magnitude is one order higher then the other maxima, which is typical for a first-order phase transition. In order to support our claim, that the peaks' positions do not significantly change above $L=48$, in Fig.~\ref{fig:c-T_L48} we also included the results obtained for $L=72$. This will also become evident later on in the respective phase diagrams in which for some selected points the phase transition temperatures estimated for $L=48$ will be compared with those determined for $L$ extrapolated to infinity. The order parameters depicted in Fig.~\ref{fig:ms1_2-T_L48} demonstrate that the transition for $D/|J|=-2$ is to the state $(\pm 1/2,\mp 1/2,0)$, characterized by a finite values of $m_{s2}$, while for $D/|J|=-1.48$ and $0$ the system tends to the state $(\pm 1/2,\pm 1/2,\mp 1)$, characterized by a finite values of $m_{s1}$. The discontinuous behavior of the order parameter for $D/|J|=-1.48$ corroborates the first-order nature of the transition.

\begin{figure}[t!]
\centering
\subfigure{\includegraphics[scale=0.5,clip]{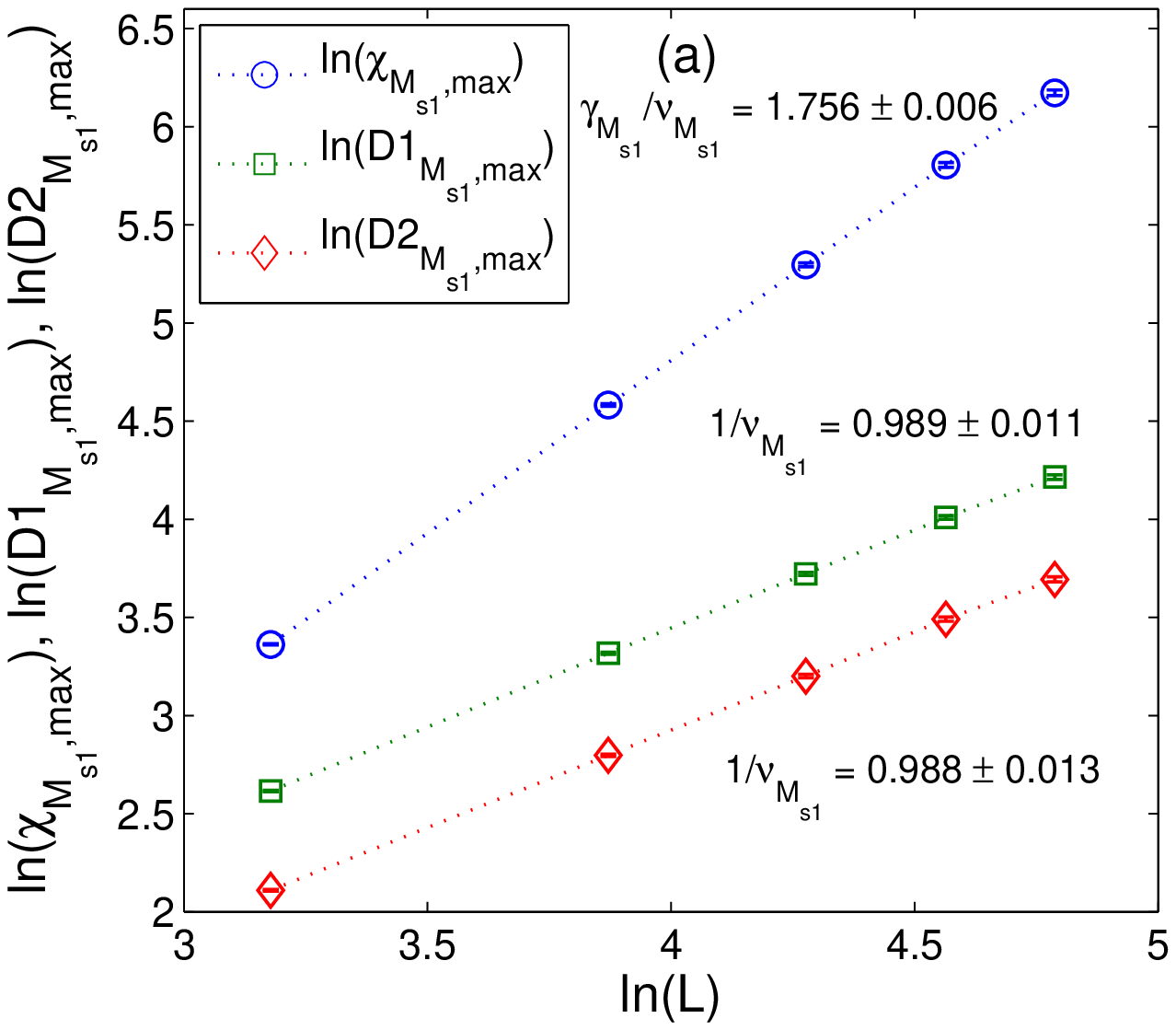}\label{fig:fss_D0}}
\subfigure{\includegraphics[scale=0.5,clip]{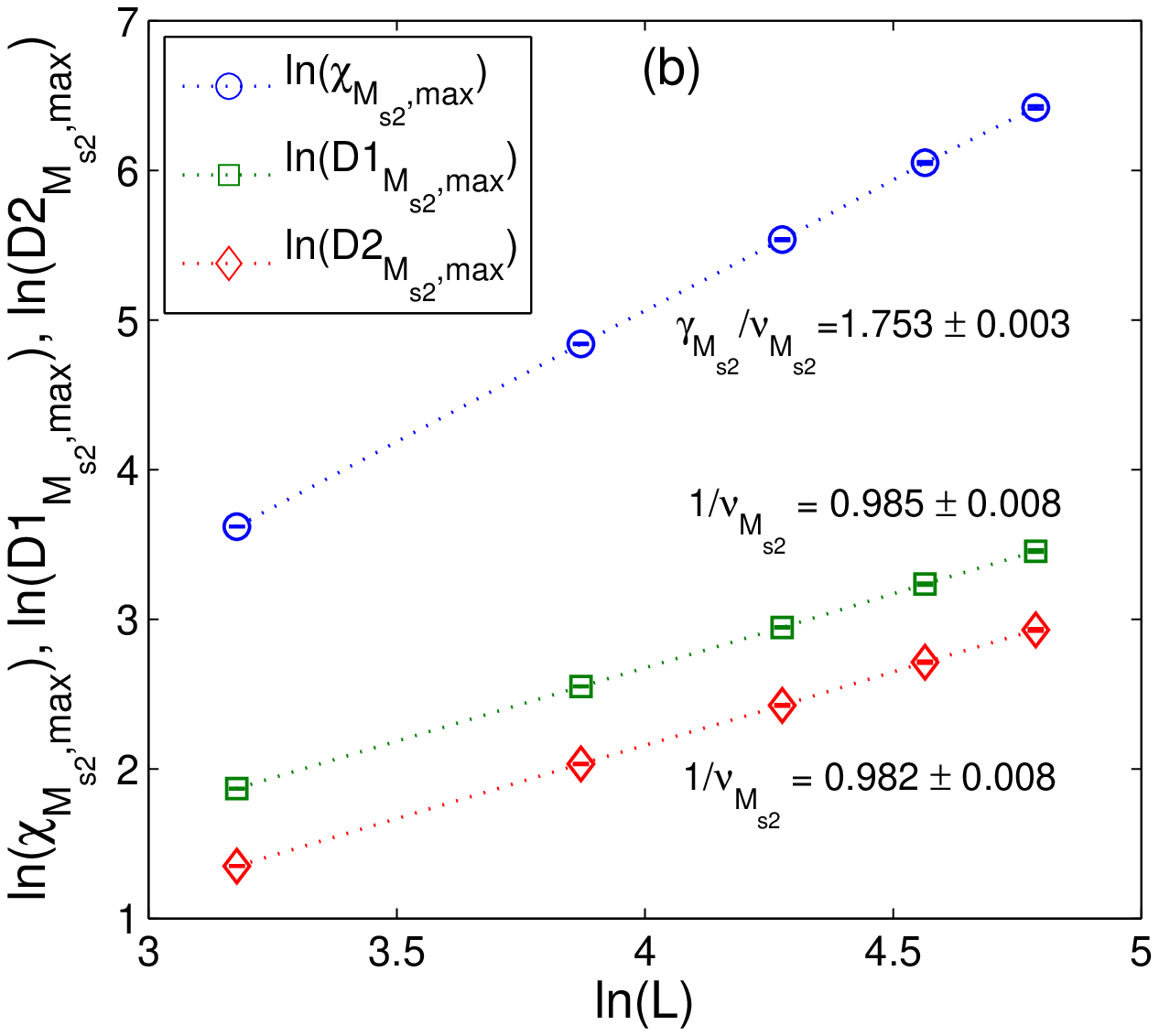}\label{fig:fss_D-2}}\vspace{-3mm}\\
\subfigure{\includegraphics[scale=0.5,clip]{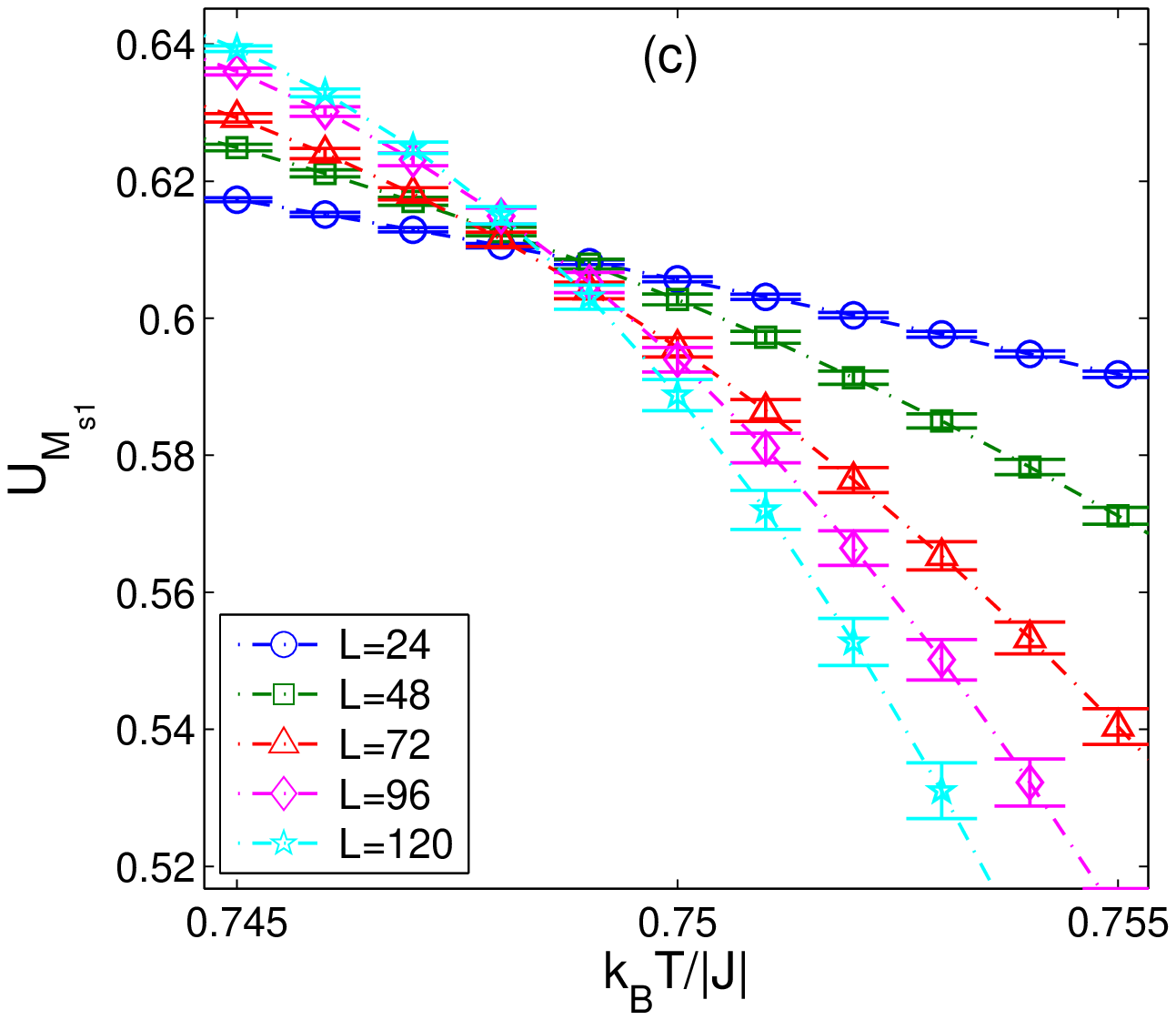}\label{fig:binder_D0}}
\subfigure{\includegraphics[scale=0.5,clip]{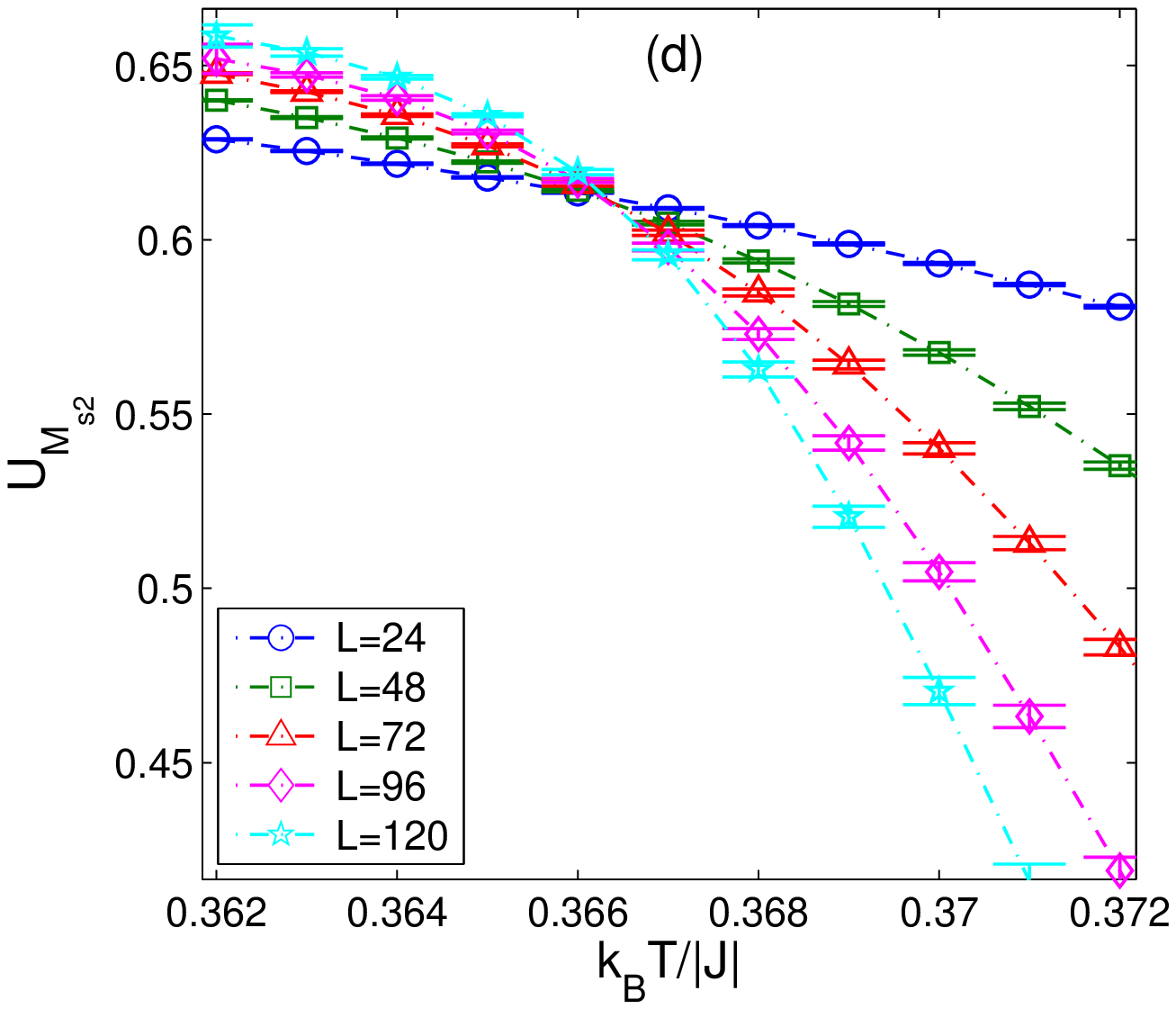}\label{fig:binder_D-2}}\vspace{-3mm}\\
\subfigure{\includegraphics[scale=0.5,clip]{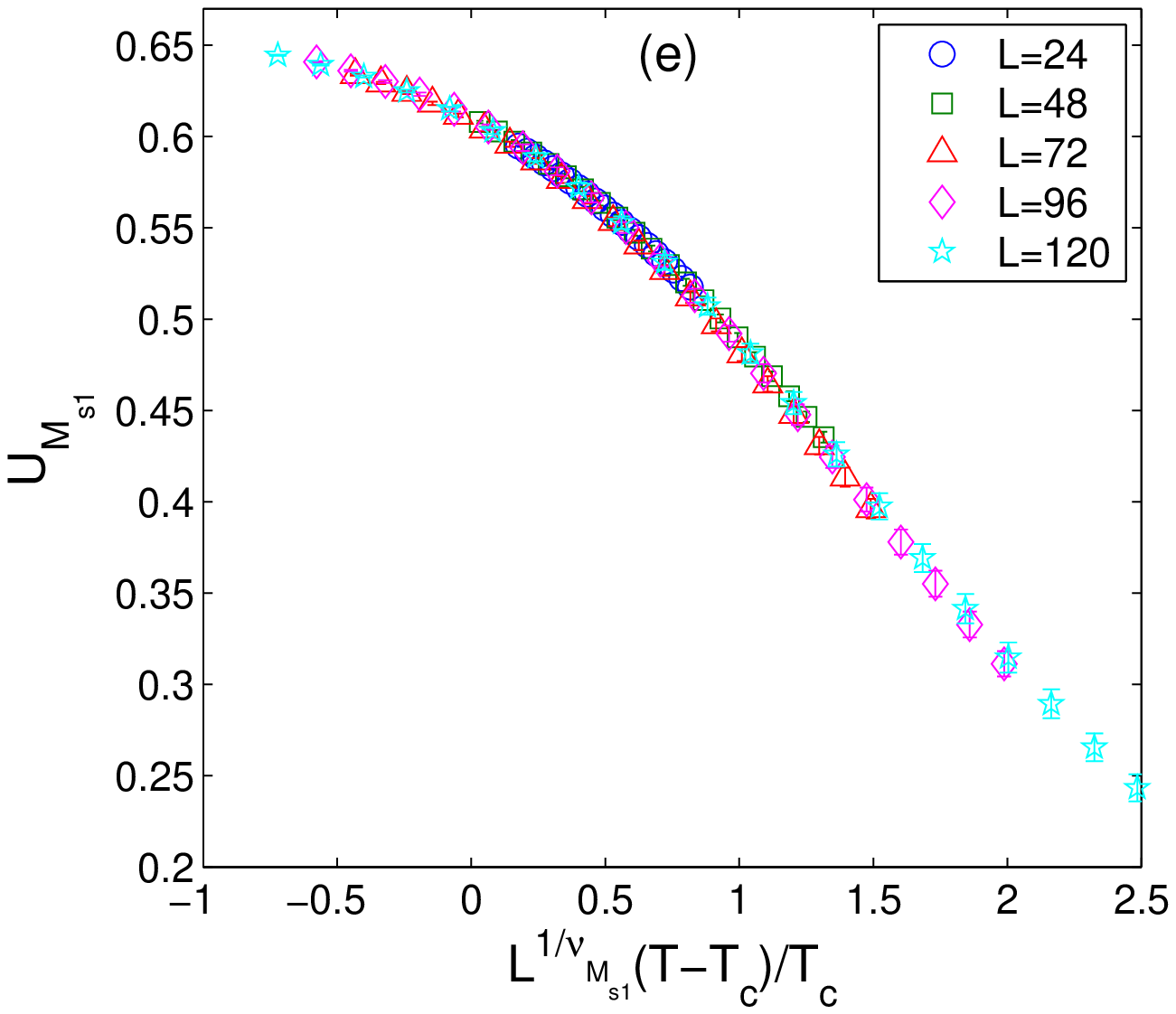}\label{fig:binder_D0_collapse}}
\subfigure{\includegraphics[scale=0.5,clip]{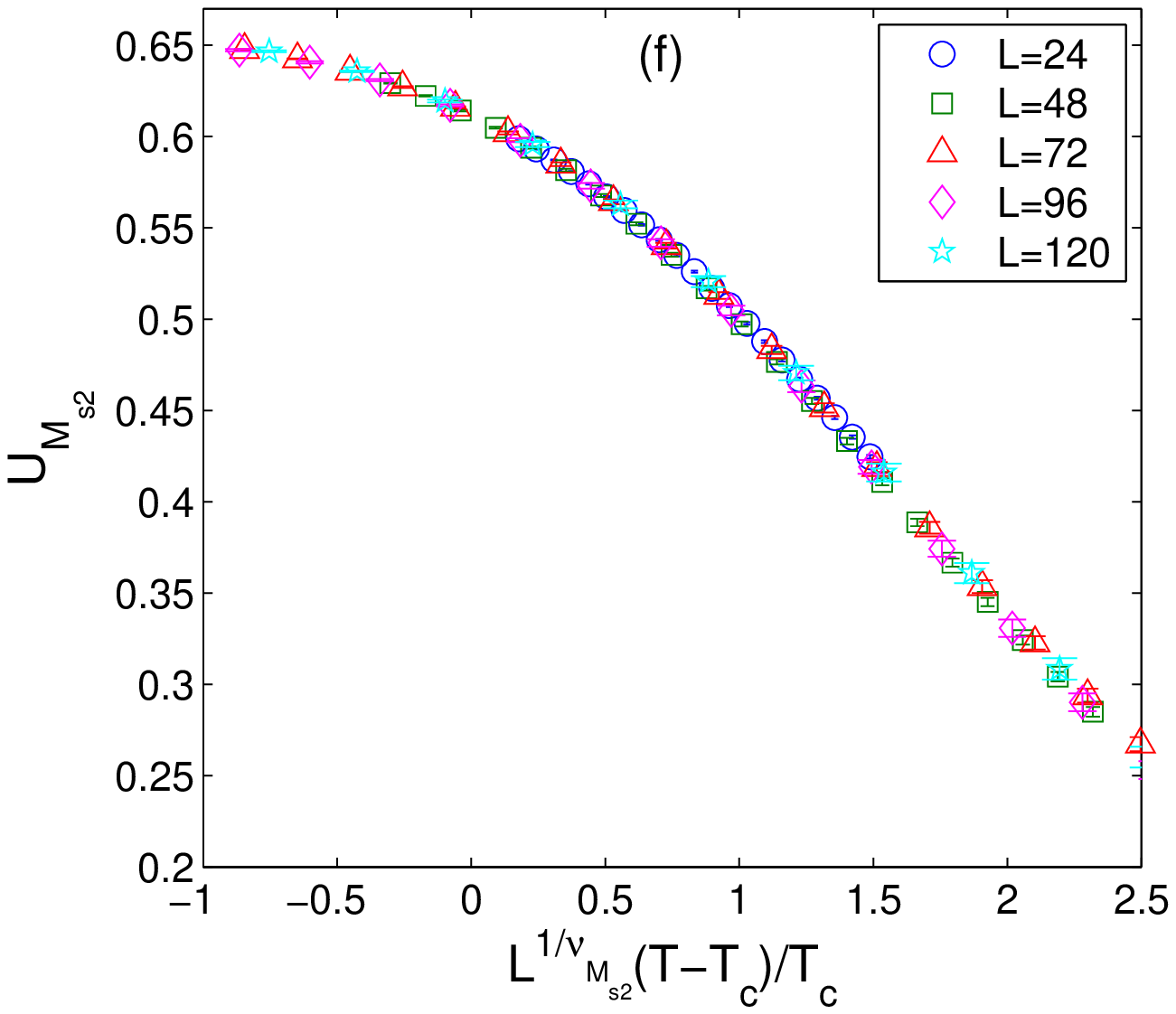}\label{fig:binder_D-2_collapse}}\vspace{-3mm}
\caption{(Color online) FSS analysis of the critical exponents ratios $1/\nu_O$ and $\gamma_{O}/\nu_O$ (a,b), the fourth-order cumulant $U_O$ temperature dependencies for different $L$ (c,d) and the $U_O$ data collapse analysis (e,f), for $O=M_{s1}$ at $D/|J|=0$ (left column) and for $O=M_{s2}$ at $D/|J|=-2$  (right column). The coefficients of determination $R^2$ for the respective fits (from top to bottom) are 0.9999, 0.9998, 0.9997 in (a) and 0.9999, 1.0000, 1.0000 in (b).}\label{fig:fss_mix1}
\end{figure}

By FSS analysis at two representative values of $D/|J|=0$ and $-2$, selected on either side of the critical value $D_c/|J|=-3/2$, we confirmed that the disorder-to-order phase transitions to both $(\pm 1/2,\pm 1/2,\mp 1)$ and $(\pm 1/2,\mp 1/2,0)$ ferrimagnetic phases are indeed second order and belong to the standard Ising universality class. The slopes of the fitted curves in Figs.~\ref{fig:fss_D0} and ~\ref{fig:fss_D-2} represent ratios of the critical exponents $1/\nu_O$ and $\gamma_O/\nu_O$, following from the scaling relations~(\ref{scalchi})-(\ref{scalD2}), where $O=M_{si}$ ($i=1,2$). We also checked that for both $D/|J|=0$ and $-2$ the specific heat maxima follow the logarithmic scaling $c_{max} = c_0 + c_1\ln(L)$, as expected for the Ising universality class in two dimensions (not shown). Having performed the FSS analysis, these two critical points can be determined with a higher accuracy from the Binder cumulant crossing method~\cite{bind81}, as an intersection of the Binder parameter $U_O$ curves for different lattice sizes $L$ and $O=M_{si}$ ($i=1,2$) (see Figs.~\ref{fig:binder_D0} and~\ref{fig:binder_D-2}). The critical temperatures were determined as $k_BT_c/|J|=0.7485 \pm 0.001$ at $D/|J|=0$ and $k_BT_c/|J|=0.3663 \pm 0.001$ at $D/|J|=-2$. Also the critical values of the Binder cumulants $U_O(T_c)=0.611$~\cite{kame93} confirm the Ising universality class. Furthermore, in Figs.~\ref{fig:binder_D0_collapse},\ref{fig:binder_D-2_collapse} we show that for the critical temperatures determined by the Binder cumulant crossing method the data for different lattice sizes indeed collapse on a single curve. The apparent first-order character of the phase transition at $D/|J|=-1.48$ will be discussed below.

\begin{figure}[t!]
\centering
\includegraphics[scale=0.57,clip]{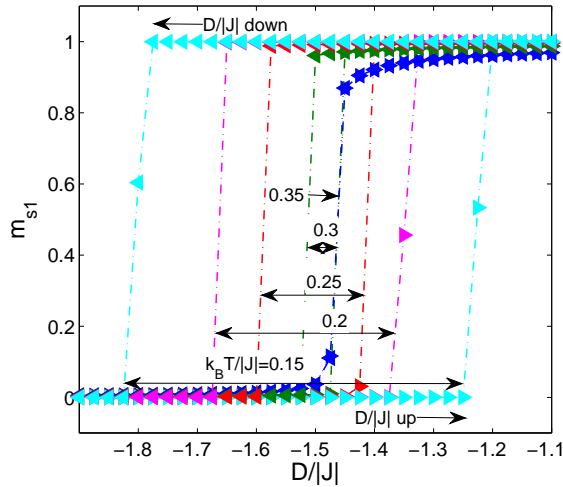}
\caption{(Color online) Order parameter $m_{s1}$ as a function of the increasing ($\triangleright$) and decreasing ($\triangleleft$) single-ion anisotropy parameter $D/|J|$ at various temperatures and $L=48$. The double-headed arrows mark the hysteresis widths.}\label{fig:hyster_L48_mix1}
\end{figure}

\begin{figure}[t!]
\centering
\subfigure{\includegraphics[scale=0.57,clip]{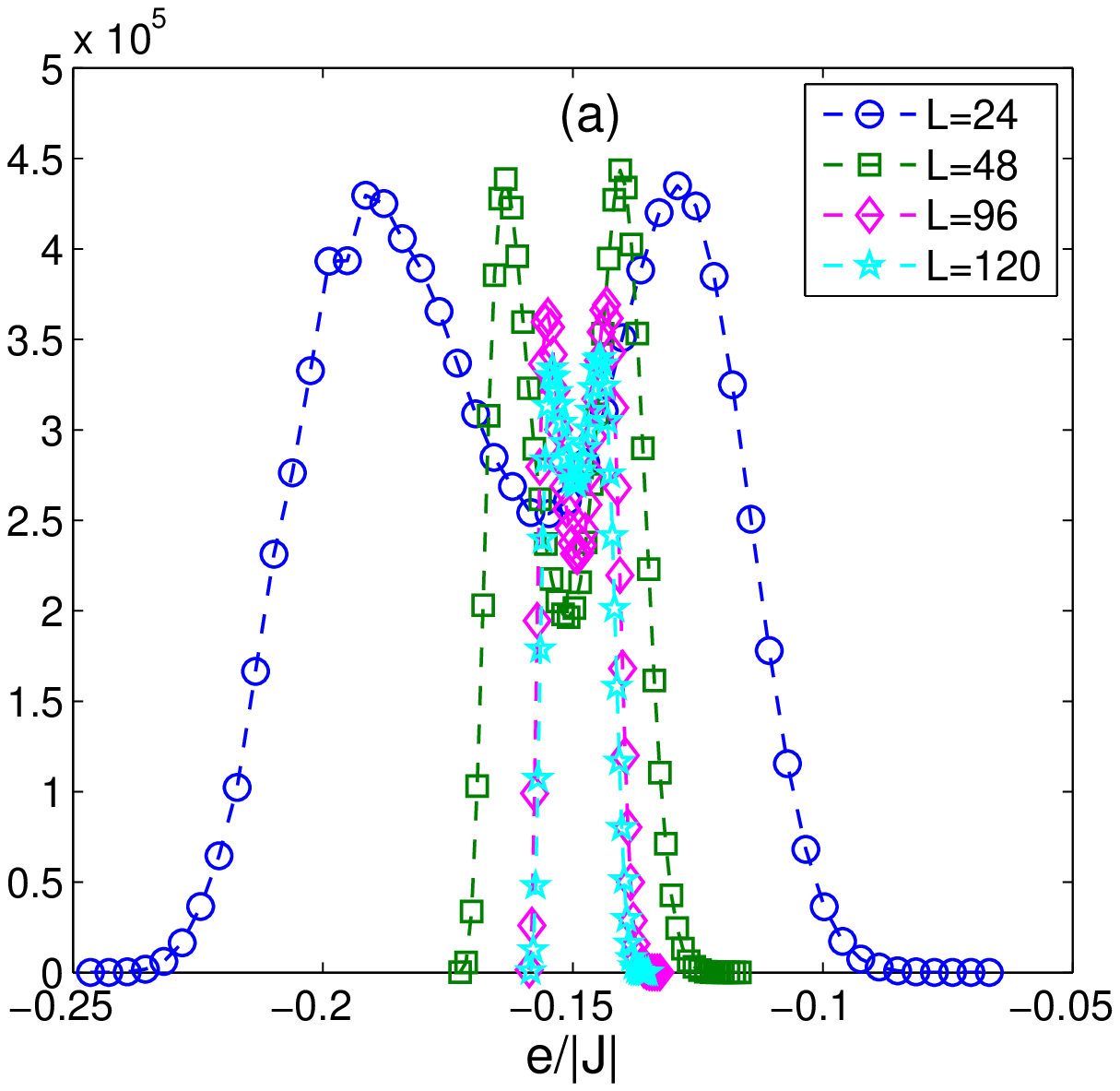}\label{fig:hist_D-1_47}}
\subfigure{\includegraphics[scale=0.57,clip]{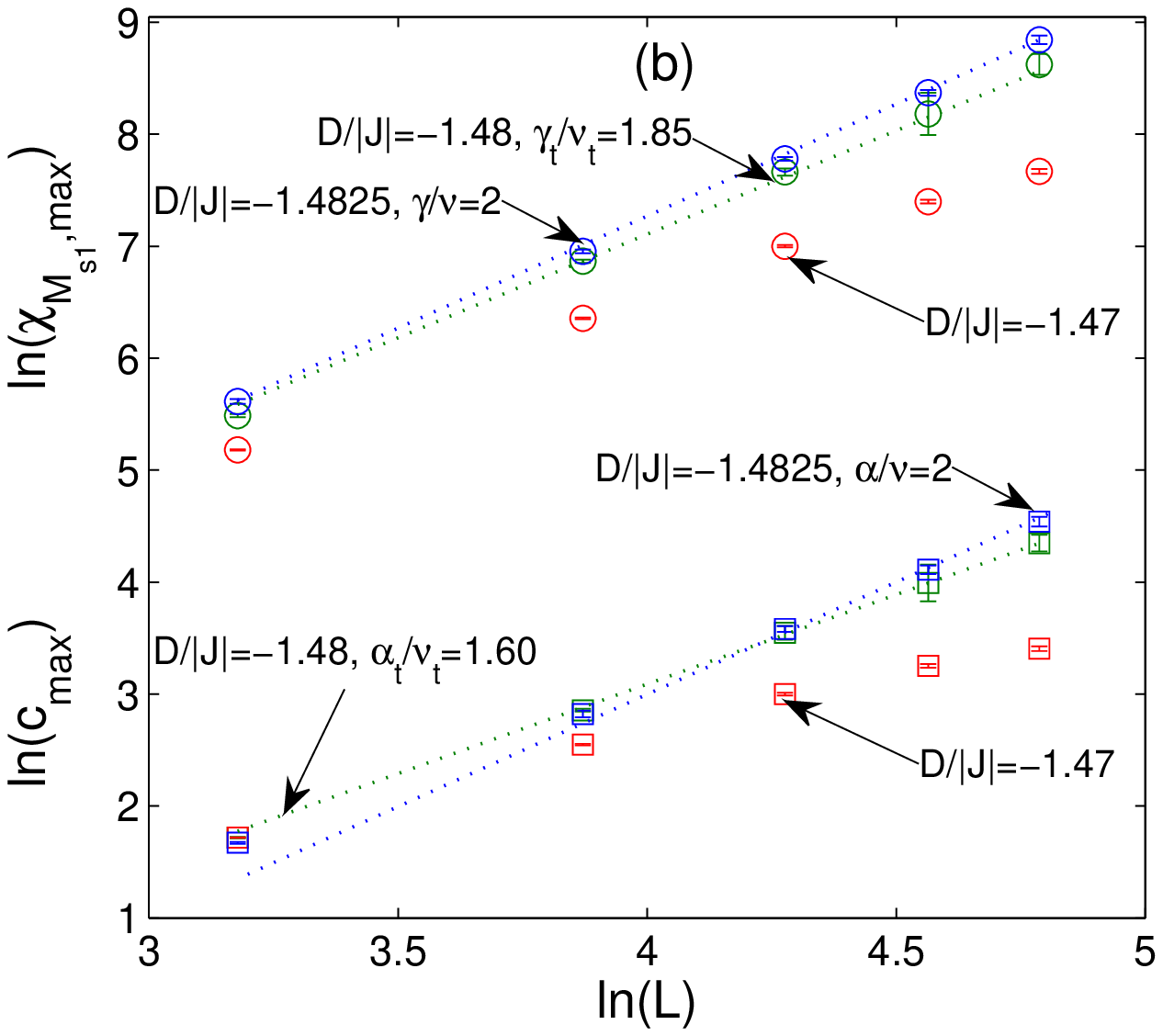}\label{fig:fss_D-1_47_-1_4825_mix1}}
\subfigure{\includegraphics[scale=0.57,clip]{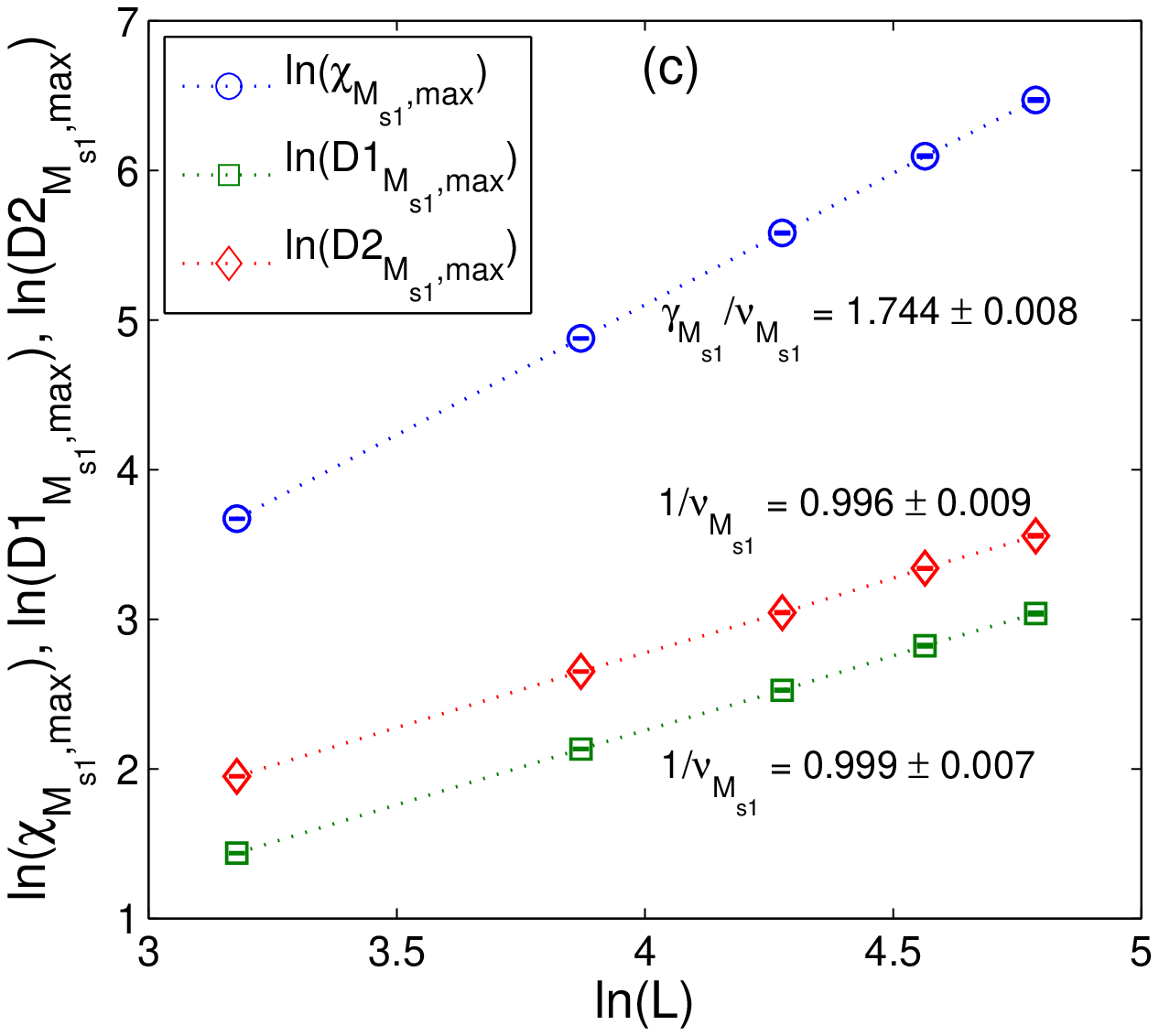}\label{fig:fss_D-1_46_mix1}}
\caption{(Color online) (a) Energy distributions for $D/|J|=-1.47$ and different $L$. The respective temperatures are tuned by the reweighing technique to achieve approximately equal peak heights. (b) FSS analysis of the susceptibility $\chi_{M_{s1}}$ and the specific heat $c$, for $D/|J|=-1.47,-1.48$ and $-1.4825$. For $D/|J|=-1.48$ and $-1.4825$ the log-log plots are respectively fitted to the exact values of the tricritical exponents ratios $\gamma_t/\nu_t=1.85$, $\alpha_t/\nu_t=1.60$ and the exponent $2$, corresponding to the system volume. (c) FSS analysis of $\chi_{M_{s1}}$, $D1_{M_{s1}}$, $D2_{M_{s1}}$ for $D/|J|=-1.46$, with the coefficients of determination $R^2$ for the respective fits (from top to bottom) 0.9999, 1.0000, 0.9999.}\label{fig:D-1_46-1_4825}
\end{figure}

Let us now examine the transition between the two low-temperature ferrimagnetic phases ${\rm FR}_1^{\rm I}$ and ${\rm FR}_2^{\rm I}$. Since the expected phase boundary is almost vertical to the x-axis, instead of the temperature dependencies of various thermodynamic functions it is more convenient to look into their single-ion parameter dependencies at a fixed temperature. By plotting the order parameters as increasing and decreasing functions of $D/|J|$ one can observe their discontinuous character and the appearance of hysteresis loops, the widths of which increase with decreasing temperature. This behavior is demonstrated in Fig.~\ref{fig:hyster_L48_mix1} for the order parameter $m_{s1}$ and $L=48$. Such behavior signals first-order phase transitions. They seem to persist even to higher temperatures at which the hysteretic behavior in not apparent any longer. The highest temperatures at which we still could observe some signs of first-order phase transitions, such as bimodal energy distribution, was $k_BT/|J|=0.35$ and the corresponding value of $D/|J|=-1.47$ (see Fig.~\ref{fig:hist_D-1_47}). 

Nevertheless, the energy barrier separating the two peaks upon initial increase seems to decrease for larger $L$, which indicate that the phase transition may not be truly first order. Indeed, when we checked whether the specific heat and susceptibility scale with the system volume, as it should be in case of a first-order transition, we found that within the used lattice sizes the linear ansatz could not be established for $D/|J|=-1.47$, as shown in Fig.~\ref{fig:fss_D-1_47_-1_4825_mix1}. We note that such a behavior that can lead to misinterpretation of a second-order transition as first order was also observed in the Blume-Capel model as well as in the frustrated $J_1-J_2$ Ising antiferromagnet on a square lattice~\cite{land81,jin13}. Clearly first-order scaling is observed only at $D/|J|=-1.4825$. Nevertheless, fairly good linear fits were also achieved in the case of $D/|J|=-1.48$ with the exponents that are between 2 and the exact tricritical values (1.85 for $\gamma/\nu$ and 1.6 for $\alpha/\nu$)~\cite{nijs79,nien79}, suggesting that the system is close to the tricritical point. On the other hand, for $D/|J|=-1.46$, the FSS presented in Fig.~\ref{fig:fss_D-1_46_mix1} clearly indicates a second-order phase transition. Thus, we roughly estimate the tricritical point at $(D_t/|J|,k_BT_t/|J|)=(-1.47 \pm 0.01,0.35 \pm 0.01)$. 

\begin{figure}[t!]
\centering
\subfigure{\includegraphics[scale=0.57,clip]{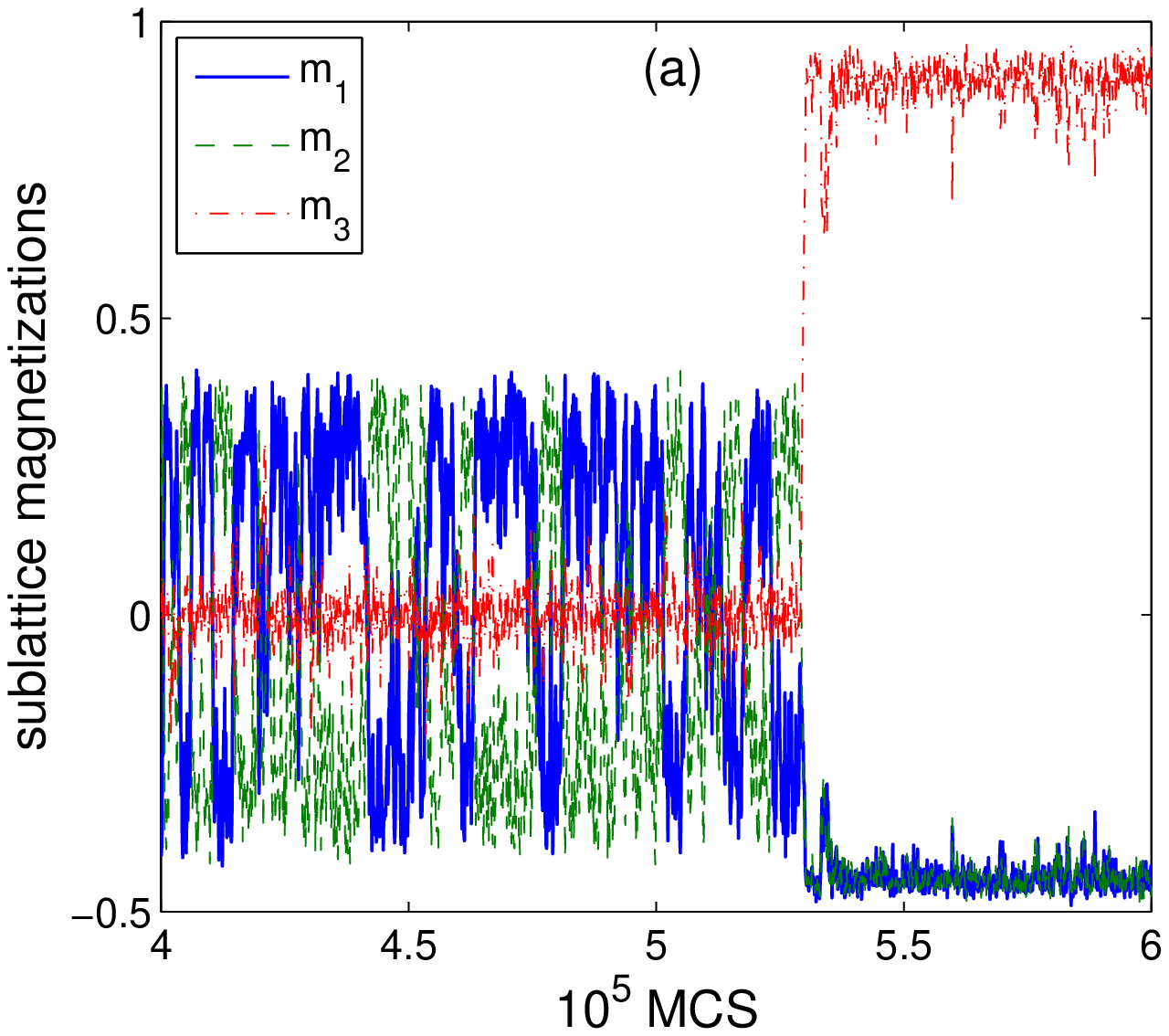}\label{fig:TS_T0_33_D-1_487_L48}}
\subfigure{\includegraphics[scale=0.57,clip]{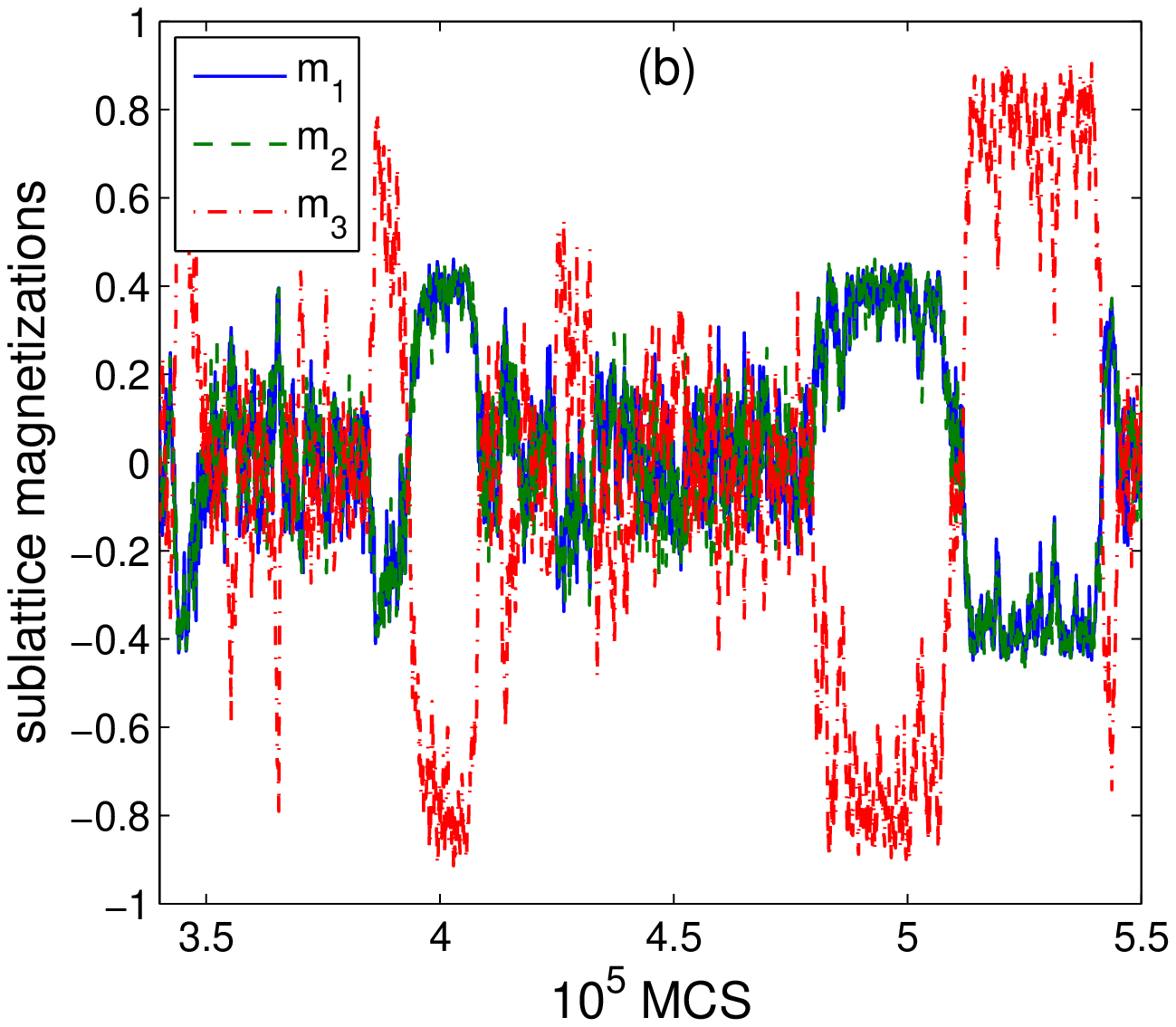}\label{fig:TS_T0_35_D-1_47_L48}}
\caption{(Color online) Time evolution of the sublattice magnetizations at the points (a) $(D/|J|,k_BT/|J|)=(-1.487,0.33)$ and (b) $(-1.47,0.35)$, for $L=48$.}\label{fig:TS_mix1}
\end{figure}

\begin{figure}[t!]
\centering
\includegraphics[scale=0.57,clip]{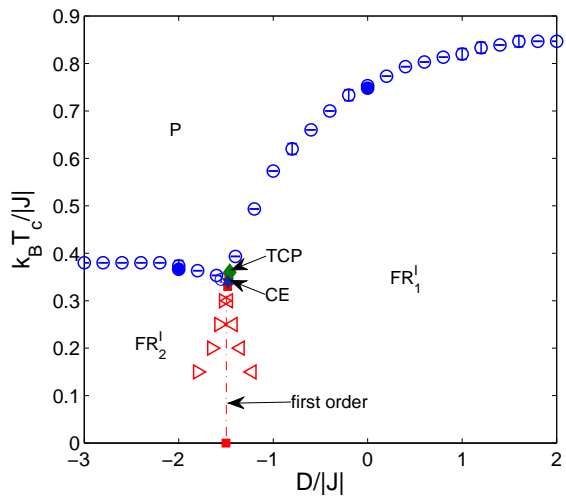}
\caption{(Color online) Phase diagram of the model I in $(k_BT/|J|-D/|J|)$ parameter space. The empty circles represent the phase transition temperatures $k_BT_c/|J|$ between the paramagnetic state P and the ferrimagnetic states ${\rm FR}_1^{\rm I}$ $(\pm 1/2,\pm 1/2,\mp 1)$ and ${\rm FR}_2^{\rm I}$ $(\pm 1/2,\mp 1/2,0)$, estimated from the specific heat peaks for $L=48$, the empty triangles mark the hysteresis widths at first-order transitions between the phases ${\rm FR}_1^{\rm I}$ and ${\rm FR}_2^{\rm I}$ with the expected phase transition boundary marked by the dash-dot line. The filled symbols at finite temperatures show more precise values obtained from the FSS analysis and the Binder cumulant crossing, where the diamond is the tricritical point (TCP), the hexagon is the critical endpoint (CE) and the square at $(D_c/|J|,k_BT_c/|J|)=(-3/2,0)$ represents the exact value of the GS transition point.}\label{fig:PD_mix1}
\end{figure}

It is interesting to notice that the phase transition at this point is no longer between the two ferrimagnetic phases ${\rm FR}_1^{\rm I}$ and ${\rm FR}_2^{\rm I}$ but between the phase ${\rm FR}_1^{\rm I}$ and the paramagnetic phase. This is in line with the above  observation of the first-order-like features of the phase transition at $D/|J|=-1.4825$ between the paramagnetic and ${\rm FR}_1^{\rm I}$ phases. The fact that there are first-order phase transitions between ${\rm FR}_1^{\rm I}$ and ${\rm FR}_2^{\rm I}$ as well as ${\rm FR}_1^{\rm I}$ and paramagnetic phases can be verified by looking at the relevant order parameters. For demonstration, in Figs.~\ref{fig:TS_mix1} we show segments of sublattice magnetization time series obtained at $(D/|J|,k_BT/|J|)=(-1.487,0.33)$ (Fig.~\ref{fig:TS_T0_33_D-1_487_L48}) and $(-1.47,0.35)$ (Fig.~\ref{fig:TS_T0_35_D-1_47_L48}). In both cases we can see discontinuous switching between two phases, however, in either point those phases are different. Namely, at $(-1.487,0.33)$ the sublattice magnetizations $(m_{\rm A}$, $m_{\rm B}$, $m_{\rm C})$ switch between the values characteristic for the states $(\pm 1/2,\mp 1/2,0)$ and $(\pm 1/2,\pm 1/2,\mp 1)$, while at $(-1.47,0.35)$ they switch between the values characteristic for the states $(0,0,0)$ and $(\pm 1/2,\pm 1/2,\mp 1)$. This finding implies the existence of a critical endpoint (CE) at which the second-order transition boundary between the paramagnetic and ${\rm FR}_2^{\rm I}$ phases joins the above discussed first-order transition line at about $(D_{ce}/|J|,k_BT_{ce}/|J|)=(-1.485 \pm 0.005,0.335 \pm 0.005)$. 

The resulting phase diagram is presented in Fig.~\ref{fig:PD_mix1}. The empty circles represent the pseudo-critical points determined from the specific heat maxima at $L=48$ and the empty triangles mark the metastable branches of the first-order phase transitions obtained from the order parameter hysteresis loops. The filled circles and squares at finite temperatures represent respectively second- and first-order transition points obtained from the FSS analysis and the filled diamond marks approximate location of the tricritical point. The filled square at $(D_c/|J|,k_BT_c/|J|)=(-3/2,0)$ shows the exact locations of the ground-state phase transition. The expected first-order phase transition boundary, marked by the dash-dot line, is obtained by a simple linear interpolation between the estimated tricritical and the exact GS transition points and only serves as a guide to the eye. We note that in this highly suppressed mixed-phase region the first-order phase boundaries can be located quite precisely, for example, by multicanonical MC simulations~\cite{berg91,zier15}. However, in our case, we are only interested in approximate location of the phase boundaries. Then, considering the exact value of the GS transition point, the estimate location of the tricritical point and assuming no anomalous behavior, such as reentrance, the low-temperature part of the phase boundary must be practically vertical to the $D/|J|$ axis. 

\subsubsection{Model II: ${\bf S}=(1/2,1,1)$}

\begin{figure}[t!]
\centering
\subfigure{\includegraphics[scale=0.57,clip]{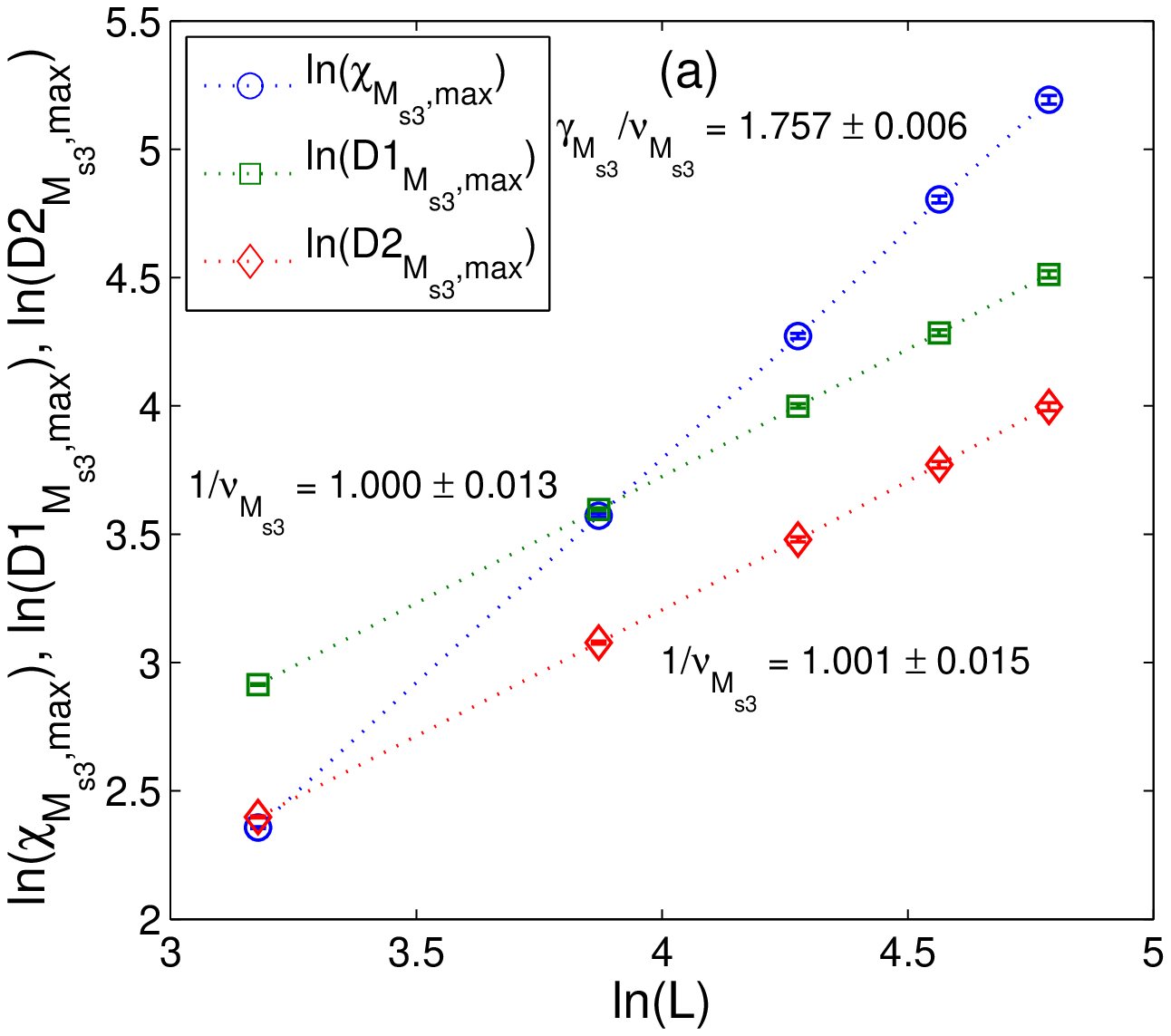}\label{fig:fss_D0_mix2}}
\subfigure{\includegraphics[scale=0.57,clip]{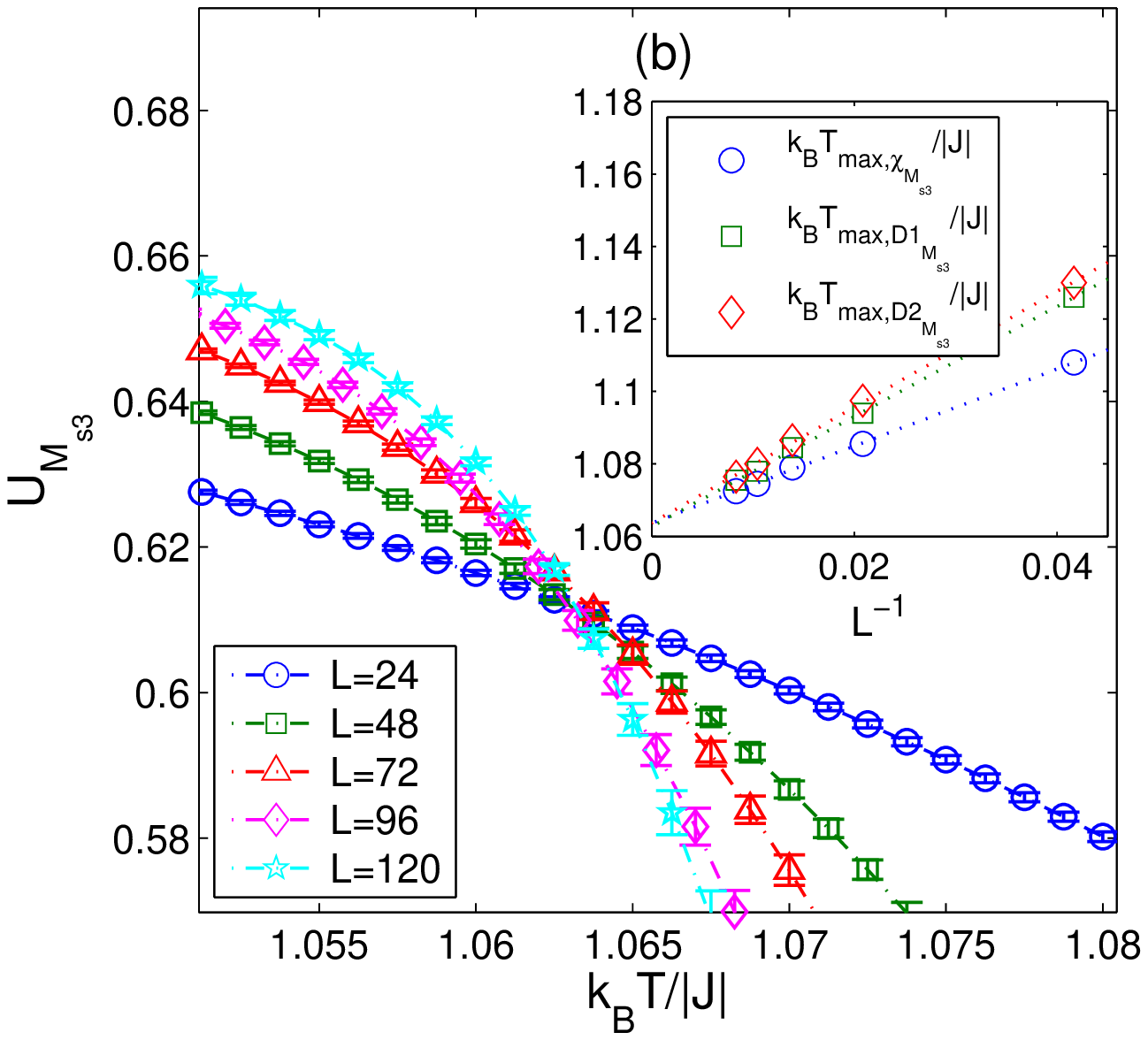}\label{fig:U_D0_L48_mix2}}
\subfigure{\includegraphics[scale=0.57,clip]{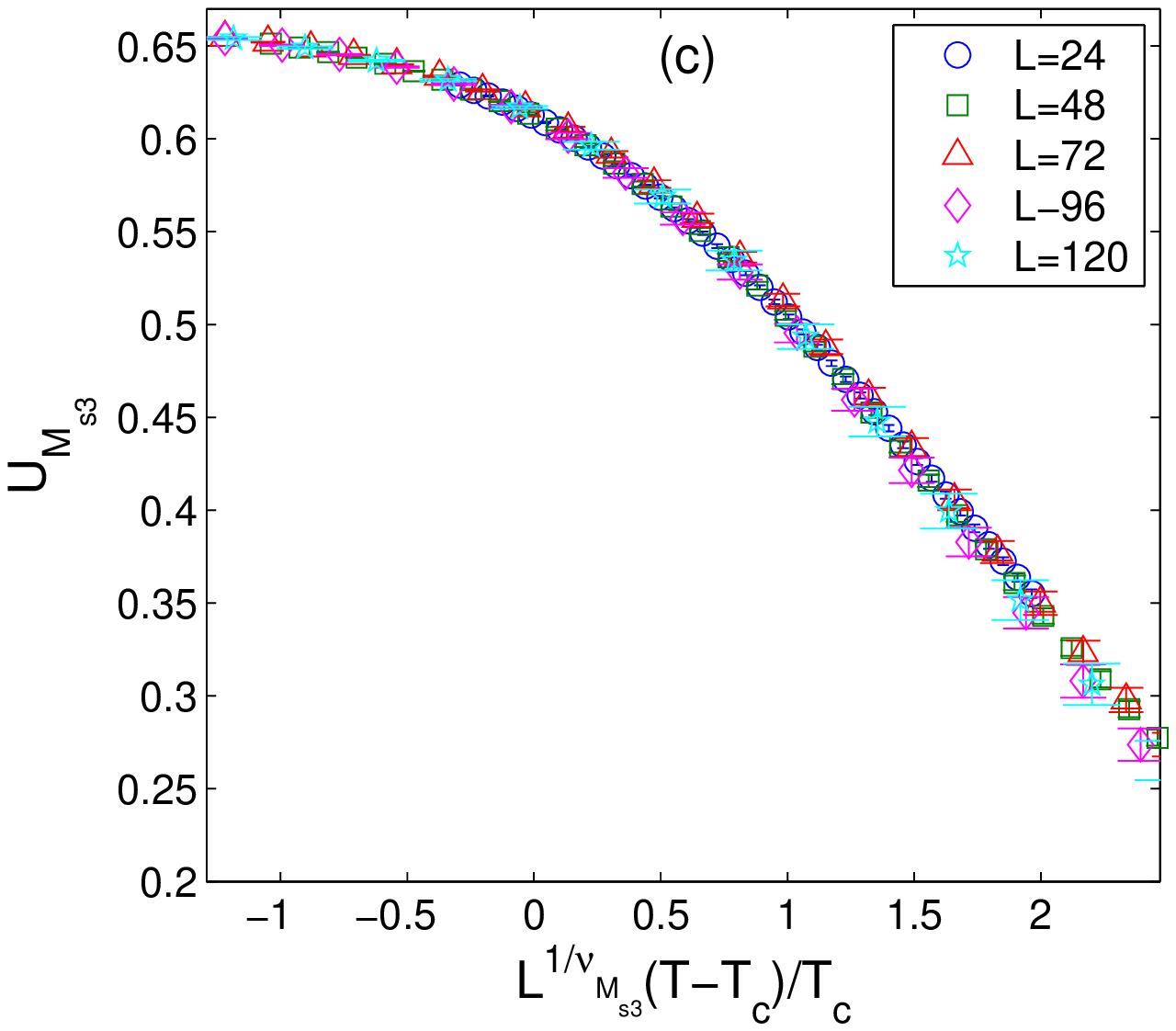}\label{fig:binder_D0_collapse_mix2}}
\caption{(Color online) (a) FSS analysis of the critical exponent ratios $1/\nu_{M_{s3}}$ and $\gamma_{M_{s3}}/\nu_{M_{s3}}$, (b) the fourth-order cumulant $U_{M_{s3}}$ temperature dependencies for different $L$, and (c) the $U_{M_{s3}}$ data collapse analysis at $D/|J|=0$. In (a) the coefficients of determination $R^2$ for the respective fits (from top to bottom) are 0.9998, 0.9999 and 0.9999. Inset in (b) shows an alternative way of the critical temperature estimation from the FSS analysis.}\label{fig:fss_mix2}
\end{figure}

Also for the model II the phase boundary as a function of the single-ion anisotropy parameter $D/|J|$ is estimated from the specific heat maxima for $L=48$, except for $D/|J|$ close to the critical value of $D_c/|J|=-3/2$, where the boundary becomes almost vertical. The only identified LRO phase is the ferrimagnetic phase ${\rm FR}^{\rm II}$: $(\pm 1/2,\pm 1,\mp 1)$, present for $D>D_c$, and the phase transition is second order complying with the standard Ising universality class. This is illustrated in Fig.~\ref{fig:fss_D0_mix2} for a selected value of $D/|J|=0$, in which we show that the obtained ratios of the critical exponents $1/\nu_{M_{s3}}$ and $\gamma_{M_{s3}}/\nu_{M_{s3}}$ are in a good agreement with the 2D Ising universality class values $1$ and $7/4$, respectively. We also checked the consistency of the specific heat critical exponent value $\alpha=0$ by verifying the logarithmic scaling (not shown). The critical temperature for $D/|J|=0$ estimated by the Binder cumulant method (Fig.~\ref{fig:U_D0_L48_mix2}) and FSS analysis (inset in Fig.~\ref{fig:U_D0_L48_mix2}) takes the value $k_BT_c/|J|=1.0635 \pm 0.0015$ and the cumulant curves for different $L$ intersect at the universal value of $U_{M_{s3}}(T_c)=0.611$.

\begin{figure}[t!]
\centering
\includegraphics[scale=0.57,clip]{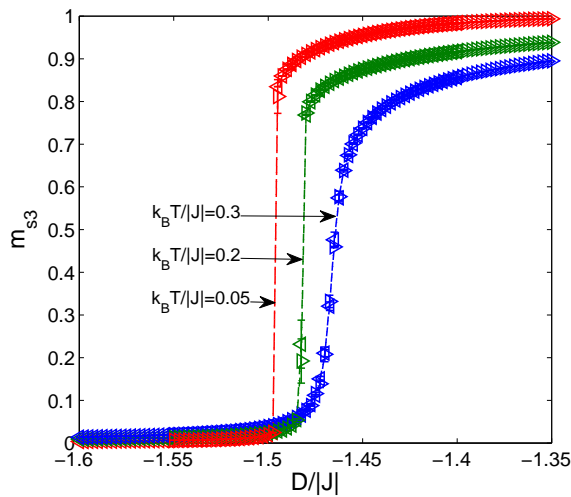}
\caption{(Color online) Order parameter $m_{s3}$ as a function of the increasing ($\triangleright$) and decreasing ($\triangleleft$) single-ion anisotropy parameter $D/|J|$ at various temperatures and $L=48$.}\label{fig:hyster_L48_mix2}
\end{figure}

\begin{figure}[t!]
\centering
\subfigure{\includegraphics[scale=0.57,clip]{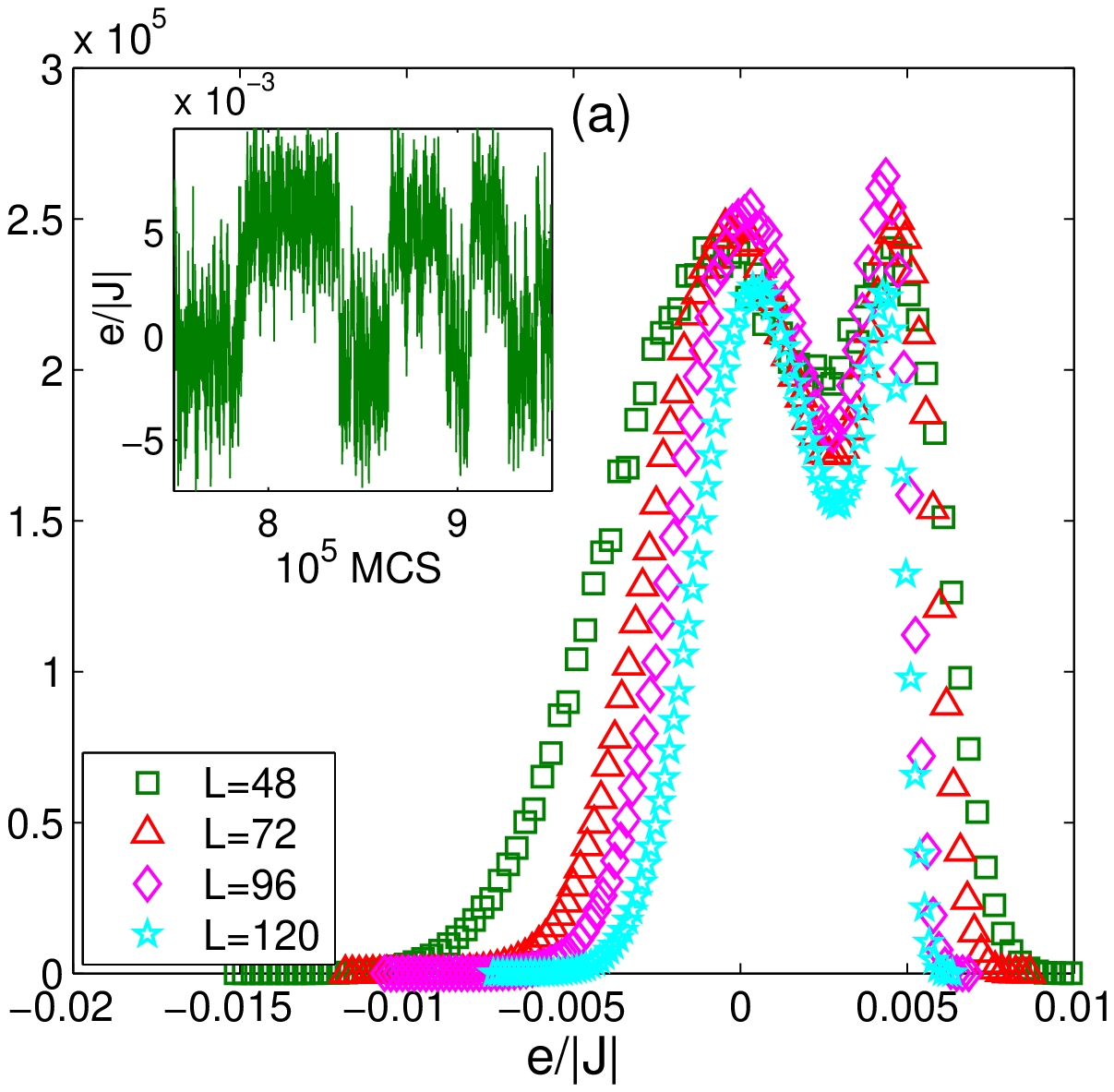}\label{fig:hist_D-1_48_mix2}}
\subfigure{\includegraphics[scale=0.57,clip]{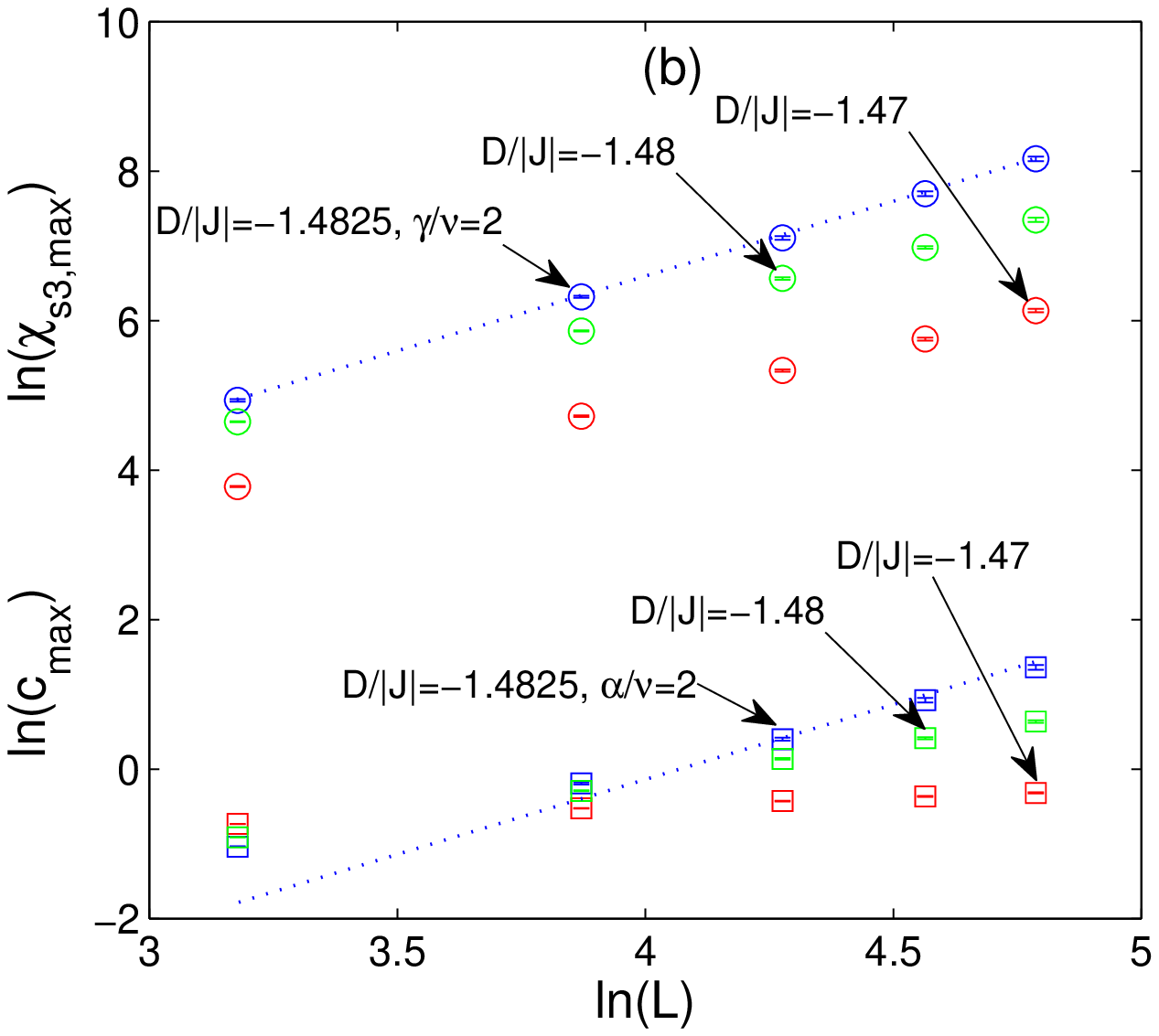}\label{fig:fss_D-1_47_-1_4825_mix2}}
\subfigure{\includegraphics[scale=0.57,clip]{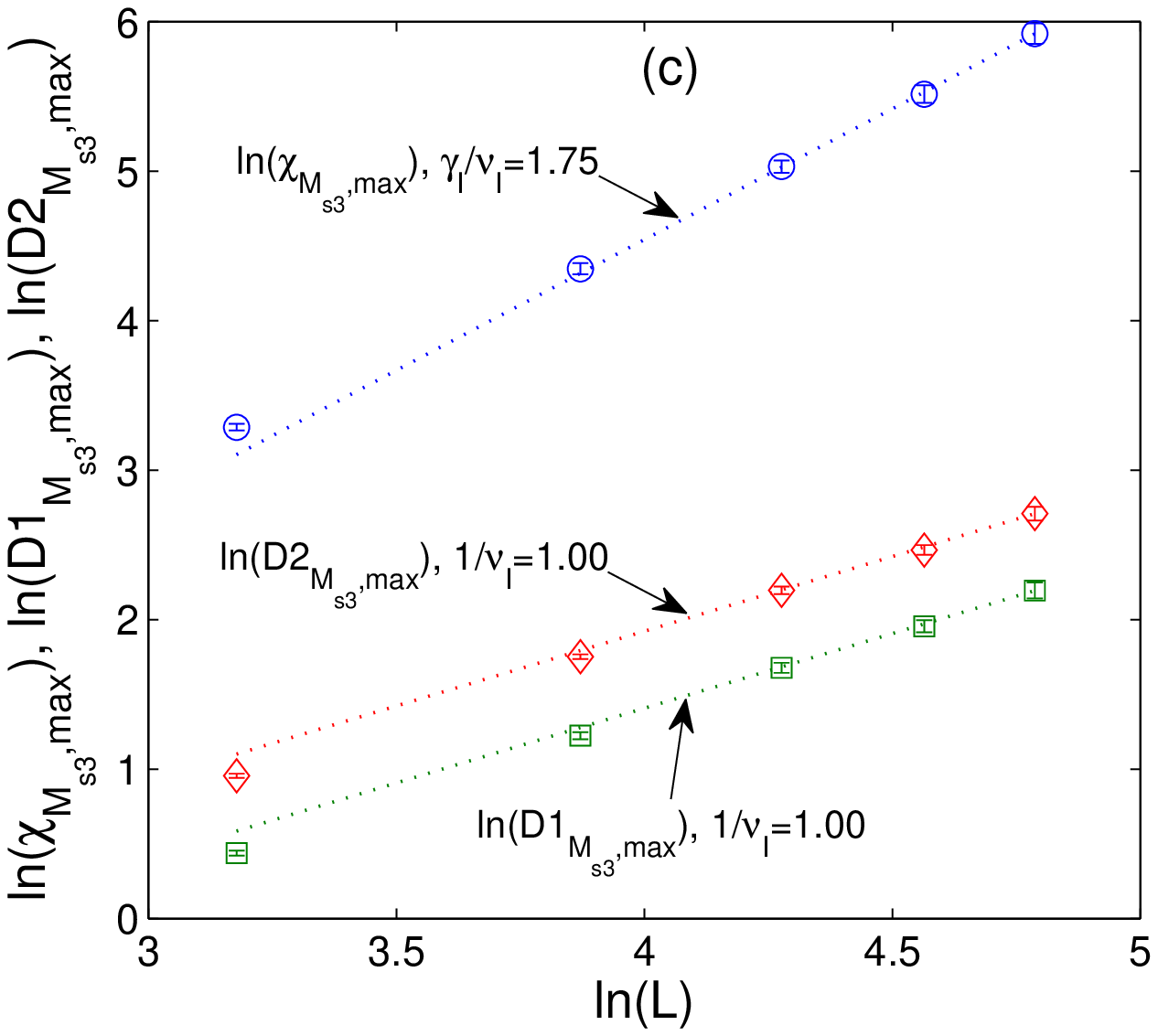}\label{fig:fss_D-1_46_mix2}}
\caption{(Color online) (a) Energy distributions for $D/|J|=-1.48$ and different $L$. The respective temperatures are tuned by the reweighing technique to achieve approximately equal peak heights. Time evolution of the internal energy shown in the inset demonstrates tunneling between the coexisting phases for $L=48$. (b) FSS analysis of the susceptibility $\chi_{M_{s3}}$ and the specific heat $c$, for $D/|J|=-1.47,-1.48$ and $-1.4825$. For $D/|J|=-1.4825$ the log-log plots are fitted to the exponent $2$, corresponding to the system volume. (c) FSS analysis of $\chi_{M_{s3}}$, $D1_{M_{s3}}$, $D2_{M_{s3}}$ for $D/|J|=-1.46$ with the fitted Ising values of the critical exponents ratios $\gamma_I/\nu_I=1.75, 1/\nu_I=1.00$.}\label{fig:D-1_4825-1_46_mix2}
\end{figure}

On approach to the critical value $D_c/|J|=-3/2$ the phase boundary rapidly drops and becomes almost vertical. Therefore, in order to locate the critical temperatures in this region, it is more convenient to measure the physical quantities at a fixed temperature as functions of the parameter $D/|J|$. At sufficiently low temperatures the measured quantities show some properties typical for first-order phase transitions. Namely, as the anisotropy parameter $D/|J|$ is decreased and increased at the fixed temperature the sublattice magnetizations, the order parameter $m_{s3}$ and the internal energy show discontinuities at some values of $D/|J|$, as demonstrated in Fig.~\ref{fig:hyster_L48_mix2} for $m_{s3}$. Nevertheless, it is interesting to notice that, in contrast to the strongly hysteretic behavior of the model I at the transition between the two ferrimagnetic phases, no apparent hysteresis can be observed in the present model. Discontinuous character of the transition reflected in bimodality of the relevant observables, such as the internal energy, disappears at higher temperatures but is still evident at temperatures slightly above $k_BT/|J|=0.2$. In Fig.~\ref{fig:hist_D-1_48_mix2} it is demonstrated for $D/|J|=-1.48$ and the temperatures $k_BT/|J|\approx 0.215$ tuned by the reweighing technique for each $L$ to achieve distributions with the two modes of about the same heights. The inset gives us some idea about the characteristic tunneling times between the coexisting phases for $L=48$. However, similar to the situation in the model I presented in Fig.~\ref{fig:D-1_46-1_4825}, for $D/|J|=-1.48$ the specific heat and staggered susceptibility maxima do not scale with volume. Such a scaling is observed only at slightly lower temperatures for $D/|J|=-1.4825$ (at least for $L \geq 72$), as shown in Fig.~\ref{fig:fss_D-1_47_-1_4825_mix2}. At higher temperatures the second-order phase transition with standard Ising critical exponents is recovered for $D/|J|=-1.46$, although again larger system sizes are required to reach the linear asymptotic regime (Fig.~\ref{fig:fss_D-1_46_mix2}). Thus, the tricritical point in the model II is roughly located at $(D_t/|J|,k_BT_t/|J|)=(-1.47 \pm 0.01, 0.27 \pm 0.04)$.

\begin{figure}[t!]
\centering
\includegraphics[scale=0.57,clip]{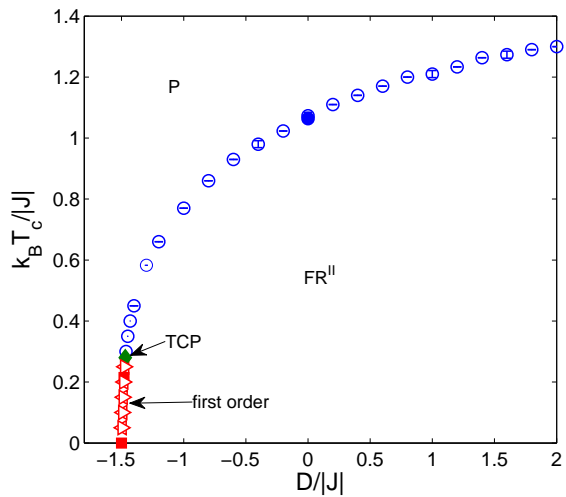}
\caption{(Color online) Phase diagram of the model II in $(k_BT/|J|-D/|J|)$ parameter space. The empty circles represent the phase transition temperatures $k_BT_c/|J|$ between the paramagnetic P and the ferrimagnetic phase ${\rm FR}^{\rm II}$ $(0,\pm 1,\mp 1)$ estimated from the specific heat peaks for $L=48$, the filled circle shows a more precise value obtained from the FSS analysis and Binder cumulant crossing, the filled diamond is the tricritical point and the filled squares represent a first-order transition point determined from FSS analysis at $D/|J|=-1.48$ and the exact value of the GS transition point at $D_c/|J|=-3/2$. The empty triangles mark the first-order transition related discontinuities in the order-parameter $m_{s3}$ in the $D/|J|$ increasing ($\triangleright$) and decreasing ($\triangleleft$) processes.}\label{fig:PD_mix2}
\end{figure}

The above results can be summarized into the phase diagram shown in Fig.~\ref{fig:PD_mix2}. As in the phase diagram of the model I, the empty circles represent the pseudo-critical points determined from the specific heat maxima at $L=48$ and the empty triangles mark the first-order phase transitions located from the jumps in the order parameter loops obtained by increasing and decreasing of the single-ion parameter $D/|J|$. The filled circle at $D/|J|=0$, the square at $D/|J|=-1.48$ and the filled diamond represent respectively second-order, first-order and tricritical points, obtained from the FSS analysis. As in Fig.~\ref{fig:PD_mix1}, the filled square at $(D_c/|J|,k_BT_c/|J|)=(-3/2,0)$ represents the exact value of the ground-state phase transition point.

\section{Conclusions}
We have studied the mixed spin-1/2 and spin-1 Ising ferrimagnets on a triangular lattice with sublattices A, B and C, in two mixing modes: $(S_{\rm A},S_{\rm B},S_{\rm C})=(1/2,1/2,1)$ (model I) and $(S_{\rm A},S_{\rm B},S_{\rm C})=(1/2,1,1)$ (model II). The pure spin-1/2 and spin-1 Ising antiferromagnets on a triangular lattice show respectively no LRO~\cite{wann50} and partial LRO for some range of a single-ion anisotropy parameter at low temperature with quasi-LRO of the Berezinskii-Kosterlitz-Thouless type at higher temperatures~\cite{zuko13}. In comparison with these models in the present ferrimagnetic models the frustration is partially accommodated by different ferrimagnetic spin arrangements and thus their critical behavior is rather different from the pure systems. On the other hand, the net magnetization of both ferrimagnetic models is always zero and thus they can show no compensation points. Therefore, from this point of view, the behavior of the present mixed-spin models is typical for antiferromagnets rather than ferrimagnets.

As for the critical properties, the model I shows two ferrimagnetic phases ${\rm FR}_1^{\rm I}$: $(\pm 1/2,\pm 1/2,\mp 1)$ and ${\rm FR}_2^{\rm I}$: $(\pm 1/2,\mp 1/2,0)$, which may be compared with the ferromagnetic case displaying two ferromagnetic phases $(\pm 1/2,\pm 1/2,\pm 1)$ and $(\pm 1/2,\pm 1/2,0)$ \cite{zuko15}. We note that on bipartite lattices the thermodynamic behavior of the systems with ferrimagnetic and ferromagnetic interactions is the same and thus the phase diagrams would be identical. However, there are substantial differences between the two phase diagrams for the triangular lattice ferromagnetic and ferrimagnetic models. First of all, due to frustration the transition temperatures for the ferrimagnetic case are significantly reduced and collapse with the ferromagnetic boundary only in the large negative $D/|J|$ limit, when the partial frustration inducing magnetic states on the C-sublattice are completely suppressed and the critical temperature for either case tends to the exact spin-1/2 Ising value on a honeycomb lattice $k_BT_c/|J|= 0.3797$~\cite{fish67}. The frustration is further increased close to the boundary between the two ferrimagnetic phases ${\rm FR}_1^{\rm I}$ and ${\rm FR}_2^{\rm I}$, which is reflected in the depression in the order-disorder phase boundary that is absent in the ferromagnetic model. Nevertheless, the frustration did not seem to affect the standard Ising values of the critical exponents. We note that besides the above presented points of $D/|J|=-2$ and 0, we also performed the FSS analysis in this region of an increased frustration at the P-${\rm FR}_1^{\rm I}$ branch of the phase diagram for $D/|J|=-1.6$ (not shown) but did not find any deviation larger than statistical errors from the standard values. However, the most conspicuous difference from the ferromagnetic case is the presence of the strongly discontinuous phase transition between the ferrimagnetic phases $(\pm 1/2,\pm 1/2,\mp 1)$ and $(\pm 1/2,\mp 1/2,0)$, which is completely absent between the ferromagnetic phases $(\pm 1/2,\pm 1/2,\pm 1)$ and $(\pm 1/2,\pm 1/2,0)$. This can be explained by the fact that in the ferrimagnetic case dramatic changes occur when the two spin-1/2 sublattices A and B, forming a connected honeycomb backbone, switch their magnetizations between $1/2$ and $-1/2$ and the spin-1 sublattice C between magnetic $\pm 1$ and nonmagnetic $0$ states. On the other hand, in the ferromagnetic case nothing happens in sublattices A and B and in sublattice C the change between magnetic $\pm 1$ and nonmagnetic $0$ states occurs only gradually in the isolated (mutually directly noninteracting) spins.

The model II has been shown to display only one ordered ferrimagnetic state ${\rm FR}^{\rm II}$: $(0,\pm 1,\mp 1)$ with no LRO on sublattice A and an antiferromagnetic LRO on the remaining sublattices B and C forming a honeycomb lattice. Similar phase occurs in the pure spin-1 triangular antiferromagnet for the single-ion anisotropy parameter $-3/2 < D/|J| < 0$, however, it is destabilized for $D/|J|>0$~\cite{zuko13}. On the other hand, in the present model II the increasing $D/|J|$ stabilizes the ferrimagnetic phase ${\rm FR}^{\rm II}$ and the critical temperature for $D/|J| \to \infty$ tends to $k_BT_c/|J|= 1.5188$~\cite{fish67}, i.e., the exact value of the spin-1/2 Ising model on a honeycomb lattice when the spin states $\pm 1$ are considered in the Hamiltonian instead of $\pm 1/2$. In comparison with the ferromagnetic model II, again the critical temperatures are lowered due to frustration but qualitatively the phase diagrams look similar. As the parameter $D/|J|$ is decreased both models show change of the phase transition nature from second to first order at a tricritical point. Nevertheless, interestingly, the strong hysteretic behavior observed in the ferromagnetic model accompanying the first-order phase transitions is not evidenced in the ferrimagnetic one. 

\begin{acknowledgments}
This work was supported by the Scientific Grant Agency of Ministry of Education of Slovak Republic (Grant Nos. 1/0234/12 and 1/0331/15). The authors acknowledge the financial support by the ERDF EU (European Union European Regional Development Fund) grant provided under the contract No. ITMS26220120047 (activity 3.2.).
\end{acknowledgments}

\end{document}